\documentclass[longauth]{aa} 
\usepackage[dvipsnames]{xcolor} 
\usepackage{graphicx,txfonts}
\usepackage[breaklinks=true]{hyperref} 
\hypersetup{colorlinks=true,citecolor=Blue}
\usepackage{natbib}
\bibpunct{(}{)}{;}{a}{}{,} % to follow the A&A style
\usepackage[english]{babel}
\usepackage{blindtext}
\usepackage{pifont}
\usepackage[normalem]{ulem}
\usepackage{comment}
\usepackage{placeins}
\newcommand{\Lk}{LkH$\alpha\,330$}

\begin{document} 

 \title{Distributions of gas and small and large grains in the \Lk\, disk trace a young planetary system\thanks{Based on observations performed with VLT/SPHERE under program ID 098.C-0760(B) and 100.C-0452(A).}}

   \author
   {P.~Pinilla\inst{1,2}, M.~Benisty\inst{3,4}, N.~T.~Kurtovic\inst{1}, J.~Bae\inst{5}, R.~Dong\inst{6}, Z.~Zhu\inst{7, 8}, S.~Andrews\inst{9}, J.~Carpenter\inst{10}, C.~Ginski\inst{11,12}, J.~Huang\inst{13,14}, A.~Isella\inst{15}, L.~P\'{e}rez\inst{16}, L.~Ricci\inst{17}, G.~Rosotti\inst{18}, M.~Villenave\inst{19}, D.~Wilner\inst{9}}
   
   \institute {Max-Planck-Institut f\"{u}r Astronomie, K\"{o}nigstuhl 17, 69117, Heidelberg, Germany. \email{pinilla@mpia.de}
   \and Mullard Space Science Laboratory, University College London, Holmbury St Mary, Dorking, Surrey RH5 6NT, UK.
   \and Univ. Grenoble Alpes, CNRS, IPAG, F-38000 Grenoble, France.
   \and Unidad Mixta Internacional Franco-Chilena de Astronom\'ia, CNRS, UMI 3386. Departamento de Astronom\'ia, Universidad de Chile, Camino El Observatorio 1515, Las Condes, Santiago, Chile
   \and Department of Astronomy, University of Florida, 316 Bryant Space Science Building, Gainesville, FL 32611, USA.
   \and Department of Physics and Astronomy, University of Victoria, Victoria, BC, V8P 1A1, Canada.
   \and Department of Physics and Astronomy, University of Nevada, Las Vegas, 4505 South Maryland Parkway, Las Vegas, NV 89154, USA.
   \and Nevada Center for Astrophysics, University of Nevada, Las Vegas, 4505 South Maryland Parkway, Las Vegas, NV 89154, USA.
   \and Center for Astrophysics \textbar\, Harvard \& Smithsonian, 60 Garden St., Cambridge, MA 02138, USA.
   \and Joint ALMA Observatory, Avenida Alonso de C\'{o}rdova 3107, Vitacura, Santiago, Chile.
   \and Anton Pannekoek Institute for Astronomy, University of Amsterdam, Science Park 904, 1098XH Amsterdam, The Netherlands.
   \and Leiden Observatory, Leiden University, 2300 RA Leiden, The Netherlands.
   \and NASA Hubble Fellowship Program Sagan Fellow
   \and Department of Astronomy, University of Michigan, 323 West Hall, 1085 S. University Avenue, Ann Arbor, MI 48109, USA.
   \and Department of Physics and Astronomy, Rice University, 6100 Main Street, MS-61, Houston, TX 77005, USA.
   \and Departamento de Astronom\'{i}a, Universidad de Chile, Camino El Observatorio 1515, Las Condes, Santiago, Chile.
   \and Department of Physics and Astronomy, California State University Northridge, 18111 Nordhoff Street, Northridge, CA 91330, USA.
   \and School of Physics and Astronomy, University of Leicester, Leicester LE1 7RH, UK.
   \and Jet Propulsion Laboratory, California Institute of Technology, 4800 Oak Grove Drive, Pasadena, CA 91109, USA.}
   \date{}

  \abstract
  {Planets that are forming around young stars are expected to leave clear imprints in the distribution of the gas and dust of their parental protoplanetary disks. In this paper, we present new scattered light and millimeter observations of the protoplanetary disk around \Lk, using SPHERE/VLT and ALMA, respectively. The scattered-light SPHERE observations reveal an asymmetric ring at around 45\,au from the star in addition to two  spiral arms with similar radial launching points at around 90\,au. The millimeter observations from ALMA  (resolution of 0.06''$\times$0.04'') mainly show an asymmetric ring located at 110\,au from the star. In addition to this asymmetry, there are two faint symmetric rings at 60\,au and 200\,au. The $^{12}$CO, $^{13}$CO, and C$^{18}$O lines seem to be less abundant in the inner disk (these observations have a resolution of 0.16''$\times$0.11''). The $^{13}$CO peaks at a location similar to the inner ring observed with SPHERE, suggesting that this line is optically thick and traces variations of disk temperature instead of gas surface-density variations, while the C$^{18}$O peaks slightly further away at around 60\,au.  We compare our observations with hydrodynamical simulations that include gas and dust evolution, and conclude that a 10\,$M_{\rm{Jup}}$ mass planet at 60\,au and in an eccentric orbit ($e=0.1$) can qualitatively explain most of the observed structures. A planet in a circular orbit leads to a much narrower concentration in the millimeter emission, while a planet in a more eccentric orbit leads to a very eccentric cavity as well. In addition, the outer spiral arm launched by the planet changes its pitch angle along the spiral due to the eccentricity and when it interacts with the vortex, potentially appearing in observations as two distinct spirals. Our observations and models show that \Lk\, is an interesting target to search for (eccentric-) planets while they are still embedded in their parental disk, making it an excellent candidate for studies on  planet-disk interaction.}


   \titlerunning{ALMA and SPHERE Observations of \Lk}
   \authorrunning{P.~Pinilla et al.}
   \maketitle

%
%________________________________________________________________
%%%%%%%%%%
\section{Introduction}                  \label{sect:intro}
%%%%%%%%%%

Most of the information that we have about planets forming in protoplanetary disks comes from observations of the dust scattering and emission. We can access the distribution of the micron-sized particles at the surface layers of the disks using optical and near-infrared scattered light observations, while the distribution of the larger particles, that is, the pebbles (millimeter- and centimeter-sized particles), is obtained from (sub-) millimeter observations. From the combination of these two techniques, it is possible to understand if the distribution of small and large particles is different in disks, which can give hints about the main mechanisms that rule the gas evolution \citep[e.g.,][]{pinilla_youdin2017}. This is because small dust grains are well coupled to the gas and follow the gas distribution, while  large particles that are partly decoupled from the gas settle to the midplane and migrate quickly inwards toward the central star, unless they are trapped in a pressure bump \citep{whipple1972}.

One of the first discoveries with the Atacama Large Millimeter/submillimeter Array (ALMA)  in the field of planet formation was the confirmation of highly asymmetric disks; for example, the disks around HD\,142527,  Oph\,IRS\,48, and HD\,135344B \citep{casassus2013, nienke2013, perez2014}.  HD\,142527 and HD\,135344B also show spiral arms in high angular resolution observations of their scattered light with SPHERE \citep[][]{avenhaus2014, stolker2016}. Other examples are V1247\,Ori and MWC\,758 \citep[][respectively]{kraus2017, dong2018}.  Only a few disks show spirals at both near-infared and submillimeter wavelengths \citep[e.g., HD\,100453 and WaOph\,6,][]{rosotti2020, brown2021, huang2018}. Interestingly,  spiral arms in scattered light are found mainly around stars toward the end of their pre-main-sequence evolution \citep[][]{garufi2018}, suggesting that the observed spiral arms  are unlikely to originate from gravitational instability, which is expected in young massive disks \citep[e.g.,][]{kratter2016}.

Potential origins of the observed asymmetries in the millimeter emission are vortices and disk eccentricity \citep[][]{ataiee2013, zhu2014, price2018, ragusa2020}. In the case of vortices, they can originate due to embedded planets in the disk perturbing the gas density and/or velocity field and triggering the Rossby-wave instability \citep[RWI;][]{lovaloce1999, li2000, lyra2009}. Similarly, the RWI can also be triggered at the edges of dead zones that are forming vortices \citep[][]{regaly2012, flock2015}. In addition, the baroclinic instability \citep[][]{klahr2003, barge2016} can also be the origin of vortices in a disk. Another potential explanation of asymmetries is dust trapping in the trailing Lagrange point of a planet that is interacting with the disk \citep{Rodenkirch2021}.

\begin{figure*}
    \centering
    \includegraphics[width=18.0cm]{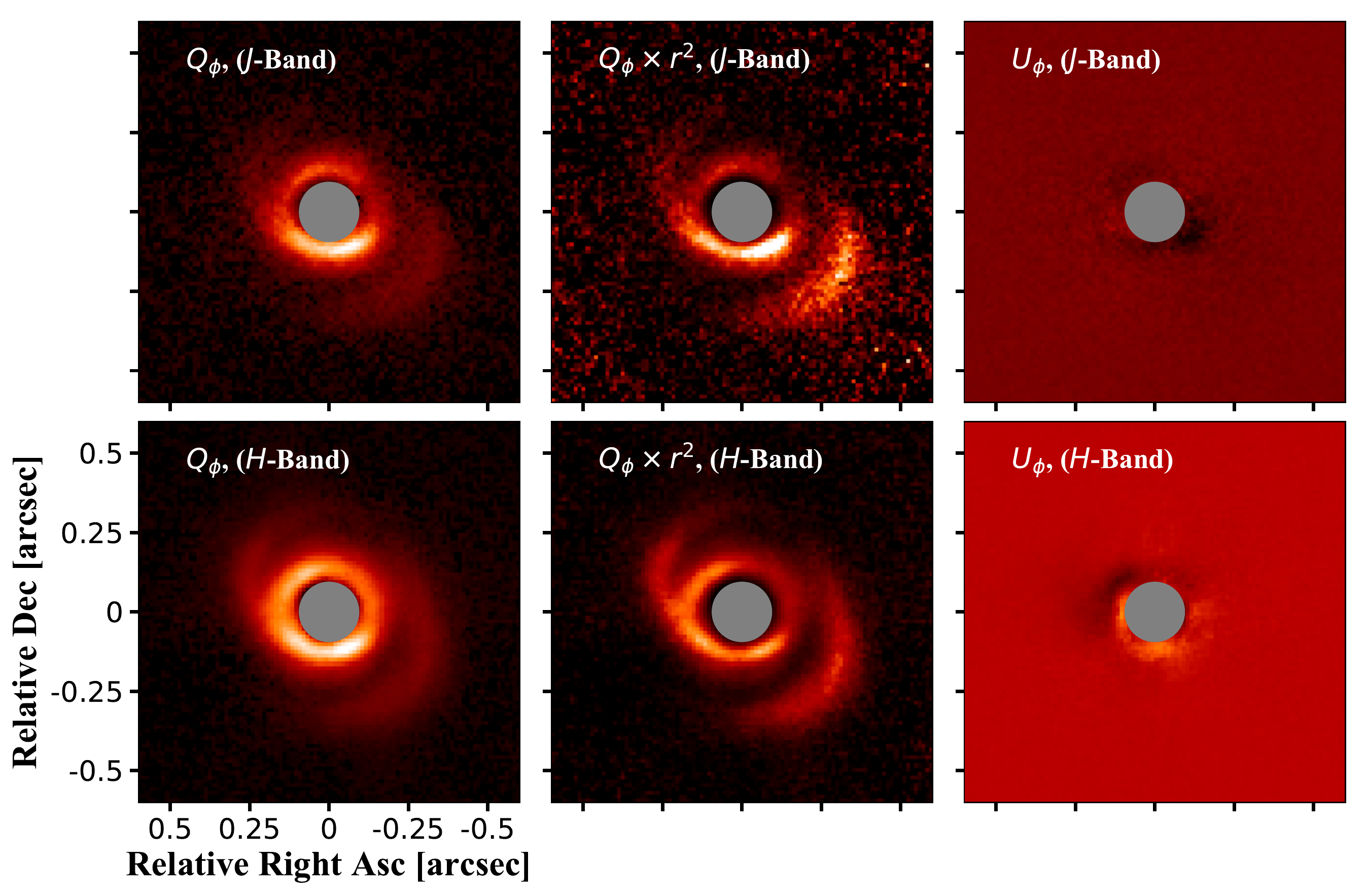}
    \caption{SPHERE observations of \Lk\, with a coronagraph of radius of 92.5\,mas in J-band (top panels) and H-band (bottom panels). From left to right: $Q_\phi$ image with a power-law normalization using an index of 0.5; $Q_\phi \times r^2$ in linear scale; and Stokes parameter $U_\phi$ also in linear scale.  In all the panels, the color scale is in arbitrary units. The coronograph is shown as a gray circle. North is up and east is left.}
    \label{fig:LkHa330_SPHERE}
\end{figure*}

Even though planets may naturally explain both spiral arms and vortices, numerical simulations have shown that the first mode of planet-driven spiral arms are usually very tight (small pitch angles) compared to the observed spiral arms in the infrared images \citep[e.g.,][]{juhasz2015, bae_zhu2018}. For this reason,  massive planets orbiting outside the spirals have been used in models to explain some of the  observed spirals in scattered light \citep[e.g.,][]{dong2015, muley2021}. 

In this paper, we present new observations from SPHERE and ALMA of the disk around \Lk, which is an  F7 star  with a stellar mass of $\sim$2.5 M$_\sun,$  a luminosity of $\sim$15 L$_\sun$ \citep[][assuming a distance of 315\,pc]{herczeg2014}, and an estimated age of $\sim$2.5\,Myr \citep[][]{uyama2018}. It is located in the Perseus molecular cloud at a distance of $\sim$318\,pc \citep[][]{gaia2016b, gaia2021edr3}. It was identified as a transition disk from observations with the Spitzer Space Telescope due to the lack of the emission in its near-infrared spectra. Observations with the Submillimeter Array (SMA) confirmed the existence of a cavity in this disk \citep[]{brown2009, andrews2011}. \cite{isella2013} combined SMA and the Combined Array for Research in Millimeter-wave Astronomy (CARMA) data at 1.3\,mm and found an asymmetric structure that potentially originated from a vortex in the disk. Recent scattered light observations of \Lk's\, disk suggested the presence of two spiral arms \citep[][]{akiyama2016, uyama2018}, which have been proposed to have originated from planet-disk interaction.

This paper is organized as follows. Section~\ref{sect:observations} summarizes the new SPHERE and ALMA observations of the disk around \Lk. Section~\ref{sect:results} describes the morphology observed with SPHERE and ALMA. Section~\ref{sect:models_obs} compares the observations with hydrodynamical simulations of gas and dust evolution in the context of planet-disk interaction, in addition to radiative transfer models being compared with observations.  Section~\ref{sect:discussion} presents the discussion about the observed structures, their origin, and the limitations of our current models to explain the observational results. Finally, Sect.~\ref{sect:conclusions} summarizes the main conclusions of this paper.

\begin{figure*}
    \centering
    \includegraphics[width=18.0cm]{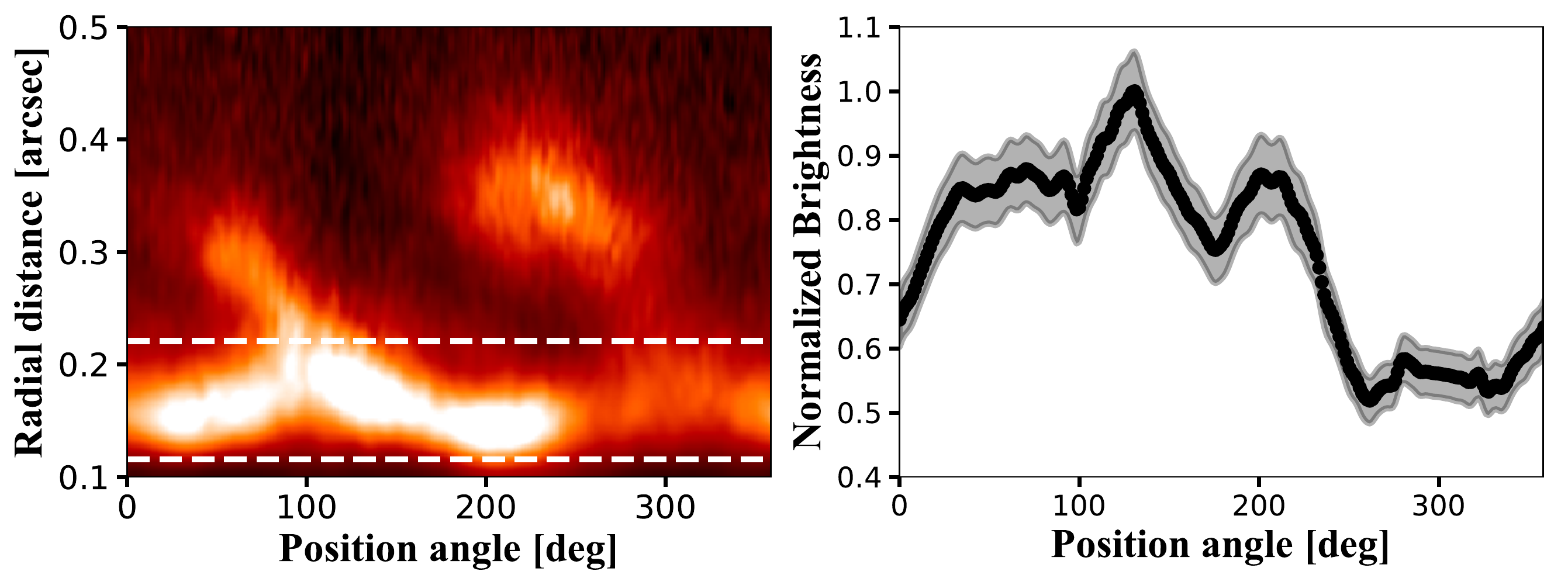}
    \caption{Polar maps of the SPHERE observations. Left panel: Polar mapping from 0.1'' to 0.5'' of the  $Q_\phi \times r^2$ image of the H-band observations after deprojection with incl=27.5$^\circ$ and PA=49.2$^\circ$. The color scale is linear, limited to 80\% of the maximum. Right panel: Azimuthal profile calculated from the mean values obtained between 0.11'' and 0.22'' (dashed lines in the left panel). The shaded areas correspond to the uncertainty of the data and come from the standard deviation in the radial and azimuth divided by the square root of the number of pixels. The data are normalized to the maximum value.}
    \label{fig:LkHa330_polar}
\end{figure*}

\section{Observations}  \label{sect:observations}

\subsection{SPHERE Observations}

We obtained observations of \Lk\  at the Very Large Telescope located at Cerro Paranal, Chile, using the SPHERE instrument \citep{beuzit2008}, a high-contrast imager with an extreme adaptive optics system \citep{sauvage2014} under program IDs 098.C-0760(B) and 100.C-0452(A) (PI: M. Benisty). In this paper, we report new polarimetric observations taken on 2017-10-05 and 2017-10-11 and obtained in the near-infrared ($J-$ (1.2\,$\mu$m) and $H-$ (1.65\,$\mu$m), respectively) with the IRDIS instrument \citep{dohlen2008}.  For all the IRDIS observations presented in this paper, we use a 185\,mas diameter coronagraph (N\_ALC\_YJH\_S) to enhance the signal-to-noise ratio in the outer disk regions. The plate scale is 12.26\,mas and 12.25 mas per pixel, for the $J-$ and $H-$ band data, respectively. 

To reduce the data, we use the public IRDAP pipeline (IRDIS Data reduction for Accurate Polarimetry) by \citet{vanHolstein2020, IRDAP}. In polarimetric differential imaging, the stellar light is split into two orthogonal polarization states, and a half-wave plate (HWP) is set to four positions shifted by 22.5$^\circ$ to construct a set of linear Stokes images. The data are then reduced following the double-difference method, from which one can derive the Stokes parameters $Q$ and $U$. If we assume single scattering events on the protoplanetary disk surface, the scattered light is linearly polarized in the azimuthal direction; therefore, we describe the polarization vector field in polar coordinates with the $Q_\phi$, $U_\phi$ Stokes images \citep{schmid2006}. In this framework, the $Q_\phi$ image contains all disk signals, while the $U_\phi$ image does not contain any. 

\begin{figure*}
    \centering
    \tabcolsep=0.05cm 
    \begin{tabular}{cc}   
        \includegraphics[width=10.5cm]{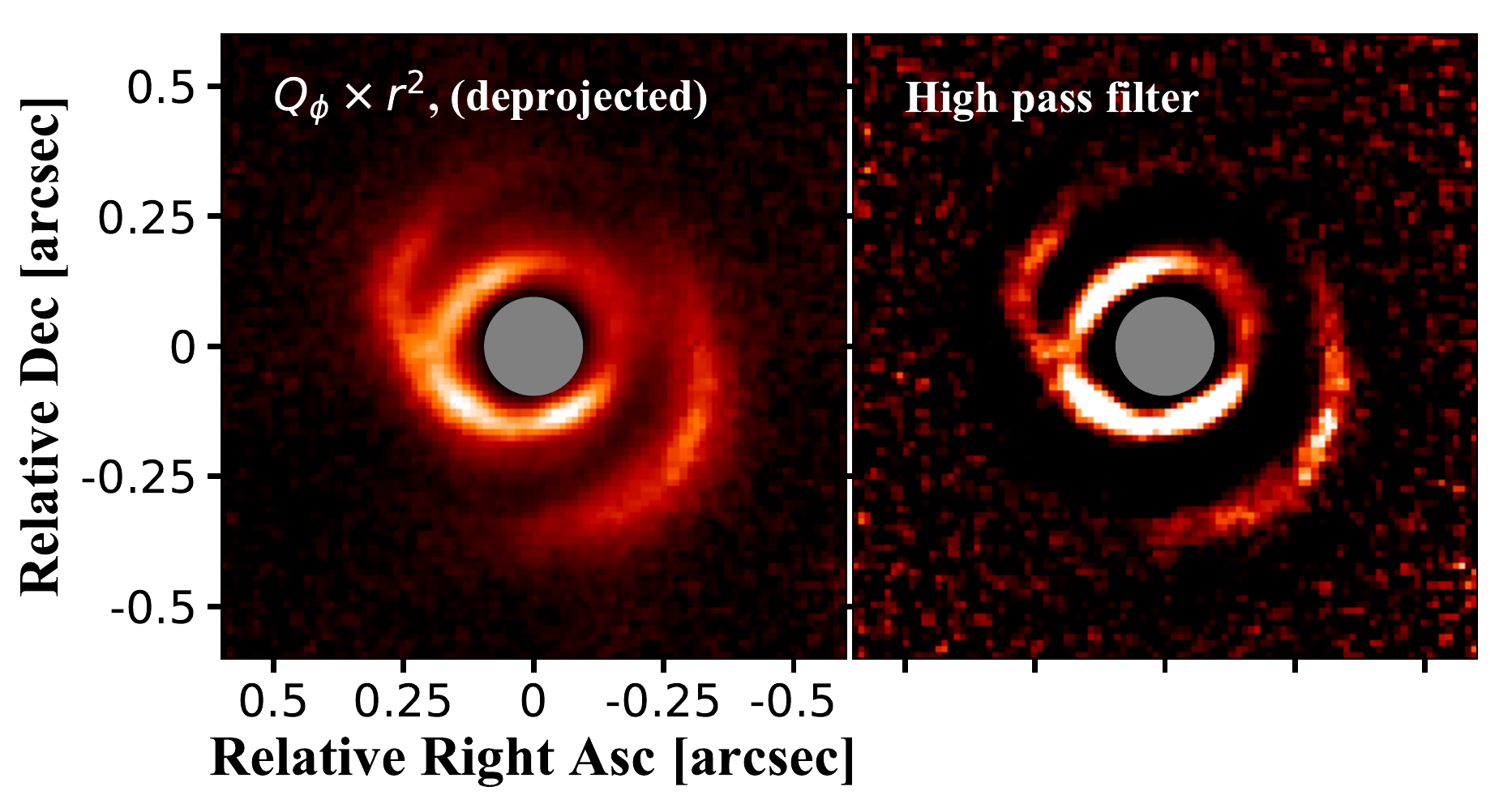}&
        \includegraphics[width=7.5cm]{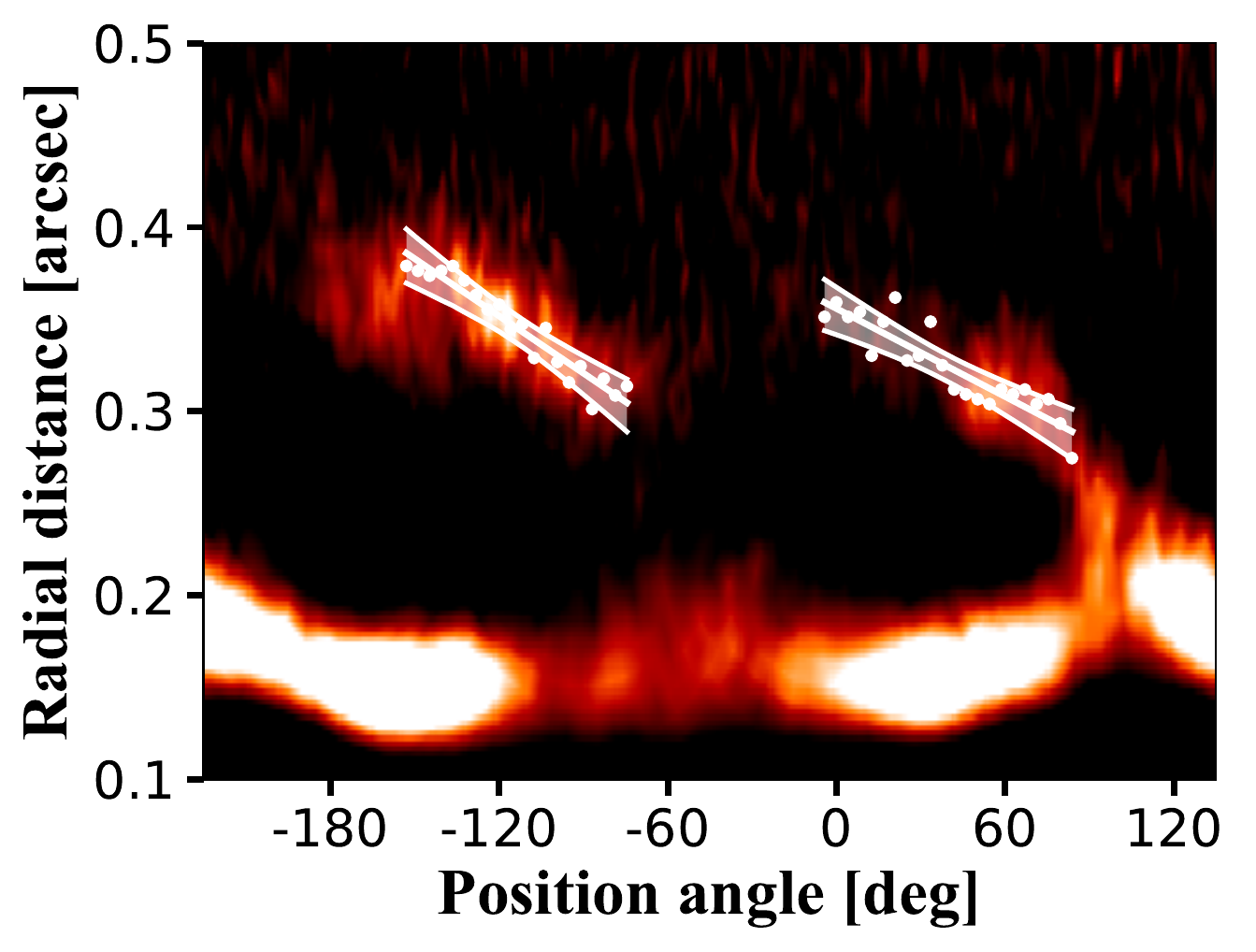}
    \end{tabular}
    \caption{SPHERE observations of \Lk. Left panel: $Q_\phi$ of the H-band observations after multiplying by $r^2$ and deprojection. Center panel: High pass filtered image. Right panel: Polar mapping from 0.1'' to 0.5'' of the sharper image in the center panel. The white lines show the spiral fit described in the text, and the shaded areas show the 1\,$\sigma$ uncertainty of the fit. In this case, the axis of the position angle is shifted in comparison with Fig.~\ref{fig:LkHa330_polar} for a better visualization of the fit.}
    \label{fig:LkHa330_deprojected_sharper}
\end{figure*}

\subsection{ALMA Observations} \label{alma_obs}

This work includes ALMA observations at 1.3\,mm (Band 6) of \Lk, which was observed on a single execution as part of the ALMA project 2018.1.01302.S (PI: M.~Benisty) on 11-Jul-2019. The correlator was configured to observe four spectral windows: two covered dust-continuum emissions were centered at $217.015\,$GHz and $233.016\,$GHz, and the two remaining ones were centered at $230.716\,$GHz to observe the molecular line $^{12}$CO ($J=2-1$), and at $219.660\,$GHz to observe the transitions $^{13}$CO ($J=2-1$) and C$^{18}$O ($J=2-1$). The channels' frequency spacing is $15.625\,$MHz for continuum and $976.562\,$kHz for the CO isotopologues lines (approximately $21\,$km\,s$^{-1}$ and $1.3\,$km\,s$^{-1}$, respectively). The total time on source is 36.79\,min, observed with 46 antennas spanning baselines from 111.2\,m to 12644.7\,m.

Using \texttt{CASA 5.6.2}, we extract the dust-continuum emission from all the windows by flagging the channels located at $\pm 25\,$km\,s$^{-1}$ from each targeted spectral line. The remaining channels from all spectral windows are averaged into 125\,MHz channels. We applied the task \texttt{statwt} to recalculate the visibilities' weight according to their observed scatter. To enhance the signal-to-noise ratio, self-calibration is applied to the data. We use a Briggs robust parameter of 0.6 for the imaging of the self-calibration process. We apply two phase calibrations and one amplitude calibration using the whole integration time as the solution interval for the amplitude calibration, and also for the first phase calibration. For the second phase calibration, we use 360s as solution interval. The overall improvement on the signal-to-noise ratio at the brightness peak is about 25\%. The calibration tables obtained from the dust-continuum self-calibration are then applied to the molecular line-emission channels. The continuum emission was subtracted using the \texttt{uvcontsub} task.

To enhance the signal-to-noise ratio in the continuum images, we apply a $uv$-tapering with a full width at half maximum (FWHM) of a 2D Gaussian  of $0.03''\times0.01''$ with position angle of $110^\circ$. The dust continuum emission image is generated using a robust parameter of 0.7, which provides us the best compromise between resolution and sensitivity. The CO lines are imaged with a robust parameter of 1.2, and a channel width of 1.5\,km\,s$^{-1}$, and an $uv$-tapering of $0.08''\times0.08''$. We find that increasing the robust value farther than 1.2 does not improve the sensitivity of the CO images, as the poor uv-coverage of our observations results in stronger side lobes of the point spread function (PSF), which are not balanced by a small increase in the beam size when going from 1.2 to 2.0 (natural weighting). The velocity width of 1.5\,km\,s$^{-1}$ is chosen to increase the sensitivity of individual channels, which were imaged with manual masking. As a final step, we apply the JvM correction to our images, which accounts for the volume ratio $\epsilon$ between the PSF of the images and the restored Gaussian of the \texttt{CLEAN} beam, as described in \citet{jorsater1995} and \citet{czekala2021}. We find $\epsilon_{\text{c}}=0.39$ and $\epsilon_{\text{l}}=0.62$ for the continuum and line images, respectively. Finally, the \texttt{bettermoments} package \citep{bettermoments} is used to create the moment maps. 

In order to reduce the data volume for the visibility analysis, we average the continuum emission into 500\,MHz width channels and 30s of time binning. We use each binned channel central frequency to convert the $uv$-coordinates into wavelength units.

\section{Results} \label{sect:results}

\subsection{Structures observed with SPHERE} \label{sect_sphere_results}

Figure~\ref{fig:LkHa330_SPHERE} shows the SPHERE ($Q_\phi$ and $U_\phi$) observations of \Lk\, in the $J-$band (1.2\,$\mu$m, top panels) and $H-$band (1.6\,$\mu$m, bottom panels). The $U_\phi$ images are almost free of any scattered light signal from the disk, and it only shows some emission in the inner ring of the H-band image.  Figure~\ref{fig:LkHa330_SPHERE} also shows the $Q_\phi \times r^2$ in linear scale to compensate for stellar illumination and better enhance the outer structures. These observations mainly reveal two types of clear structures: (a) a non-uniform brightness ring and (b) two spiral arms. 

First, the non-uniform ring is located between $0.11''$ and $0.22''$ from the star and the variation of its brightness is shown in Fig.~\ref{fig:LkHa330_polar}. The peak brightness of this ring (from the $Q_\phi$ image) is located at $0.15''$. The left panel of this figure shows the  polar mapping from $0.1''$ to $0.5''$ of the $Q_\phi \times r^2$ of the H-band image after deprojection. For the deprojection, we use  the inclination and position angle obtained from the visibility fitting to the millimeter dust-continuum emission as shown in Sect.~\ref{sect_visi_analysis} (incl=27.5$^\circ$ and PA=49.2$^\circ$).The ring shows a sinusoidal pattern in polar coordinates, and it is possible that the  deprojection  may not fully restore the “face-on” view of the disk due to the flaring of the disk surface  \citep{dong2016}. The right panel of Fig.~\ref{fig:LkHa330_polar} shows the azimuthal profile of the ring calculated from  $0.11''$ to $0.22''$, demonstrating variations of the ring brightness of $\sim 50\%$. The ring has three local brightness maxima: the main peak located at a position angle of around $\sim$130$^\circ$, and two surrounding peaks, a very wide one located at $\sim70^\circ$ and a narrower one at $\sim200^\circ$, both of them being $\sim15\%$ less bright than the main peak. The rest of the ring from $\sim250^\circ-360^\circ$ has an almost uniform brightness, which is 50\% lower than the mean peak. 

\begin{figure*}
    \centering
    \includegraphics[width=18.0cm]{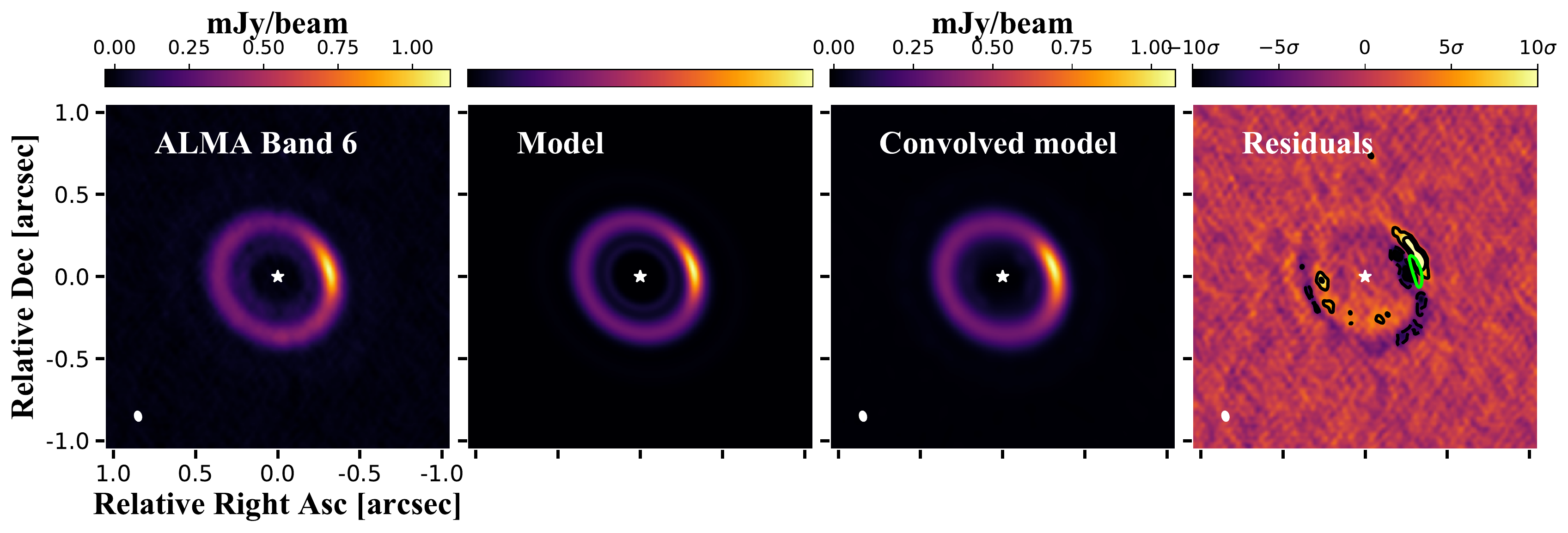}
    \caption{Observations versus model images from our best fit with \texttt{galario} (before and after convolution) and the residuals. The contours of the residual map are [$-10\sigma$, $-5\sigma$, $5\sigma$, $10\sigma$], where negative values are dashed contours and positive values solid contours. The green contour in the residual map shows the area that encloses 80\% of the peak of the emission from the observations as a reference.}
    \label{fig:ALMA_data}
\end{figure*}

The second clear set of structures are the spiral arms. To quantify the shape of the two spiral arms, we first deprojected the $Q_\phi$ of the $H-$band image (which has better signal-to-noise ratio than the $J-$band image) after multiplying by $r^2$ and took the difference between the original image and a smoother version which is obtained by convolving it with a circular Gaussian kernel ($\sigma=50\,$mas). This process is known as a high-pass filter (or unsharp masking) and it helps to sharpen the image and highlight potential small-scale structures (see Fig.~\ref{fig:LkHa330_deprojected_sharper}).  From this sharper image, we select the peak of emission along the radial direction every $4^\circ$ along the spiral arm \citep[e.g.,][]{perez2016, kurtovic2018}.

With the selected points, we calculate the pitch angle of the features by fitting an Archimedean spiral, following

\begin{equation}
    r \, = r_0 \, + b \theta \text{,}
\label{eq:archimedean_spiral}
\end{equation}

\begin{table}
\centering
\caption{Best parameters from spiral fitting following Equation \ref{eq:archimedean_spiral}. Inner and outer refer to the side of each spiral that is closer or farther from the disk center, respectively. ``mas'' stands for milliarcsecond. NE refers to the north-east, whereas SW refers to the south-west spiral.}
\begin{tabular}{ c|c|c|c } 
  \hline
  \hline
\noalign{\smallskip}
    & Spiral NE  & Spiral SW  & units \\
    &          &          &       \\
\noalign{\smallskip}
  \hline
\noalign{\smallskip}
    $r_0$       & $314.5_{-3.9}^{+3.7}$  & $419.1_{-13.5}^{+12.5}$ & mas \\
     $b$       & $1.01_{-0.11}^{+0.12}$ & $0.78_{-0.10}^{+0.10}$  & mas/deg \\
\noalign{\smallskip}
  \hline
\noalign{\smallskip}
    $r_{\text{inner}}$   & $294.4_{-5.8}^{+5.5}$     & $283.3_{-5.0}^{+5.1}$     & mas \\
    $r_{\text{outer}}$   & $378.2_{-5.3}^{+5.5}$     & $352.0_{-5.2}^{+4.9}$     & mas \\
    $\mu_{\text{inner}}$ & $11.3_{-1.4}^{+1.5}$      & $9.0_{-1.3}^{+1.3}$       & deg \\
    $\mu_{\text{outer}}$ & $8.8_{-0.9}^{+0.9}$       & $7.3_{-0.9}^{+0.8}$       & deg \\
\noalign{\smallskip}
  \hline
  \hline
\end{tabular}
\label{tab:spiral_results}  
\end{table}

\noindent where $\theta$ is the azimuthal angle, $r_0$ is the spiral position for $\theta=0$, and $b$ is a constant that relates to the pitch angle by $\mu=b/r$. We assume that both spirals share the same polar coordinate system, centered at the stellar position. We fit $(r_0, \theta)$ for each spiral with an MCMC routine based on \texttt{emcee} \citep{emcee2013}; each run has two free parameters, 128 walkers, and 1000 steps, of which the first 200 are considered burn-in steps. We minimize the $\chi^2$ between the model spiral and the measured points in the image using 1\,pix as the error for each measurement and a flat prior for both parameters.

We find that the points on the edges of the south-west feature start moving in the reversed radial direction compared to the rest of the spiral. This effect is most likely produced by deprojection effects of the flared surface layer and would require a correction with the flaring angle to be fixed \citep[][]{dong2016}. In addition, it is possible that the inclination and position angle obtained from ALMA is not exactly the same for the scattered light image since they trace different disk vertical regions. Therefore, those points that move in the reversed radial direction were not considered for the fit.

In the right panel of Fig.~\ref{fig:LkHa330_deprojected_sharper}, we show the best fit and the $3\sigma$ confidence region for each spiral, and Table~\ref{tab:spiral_results} summarizes the best parameters from this fitting. This table includes the inner and outer radii and the pitch angle, referring to the side of each spiral that is closer to or farther from the disk center. The launching point of the north-east spiral arm is 294\,mas ($\sim$94\,au), with a pitch angle of 11.3$^\circ$. The launching point of the south-west spiral arm is very similar at 283\,mas ($\sim$90\,au), with a pitch angle of 9.0$^\circ$. The farthest point of this spiral arm is at 352\,mas ($\sim$111\,au), which is similar to the location of the main asymmetry of the dust-continuum emission from ALMA (Sect.~\ref{sect_visi_analysis}). 

To test how the spiral fit is influenced by the $r^2$ scaling, we perform the same fit without this scaling. The result of this test is that the fit of north-east spiral remains nearly the same, while for the south-west spiral the points that were in a reversed radial direction at the edges of the spiral do follow the spiral without the $r^2$ scaling, which makes the pitch angle decrease by a factor of two; whereas, the launching point increases by $\sim$10 pixels ($\sim$93\,au instead of 90\,au).

\subsection{Dust and gas morphology from ALMA observations} \label{sect_visi_analysis}

\begin{figure}
    \centering
        \includegraphics[width=9cm]{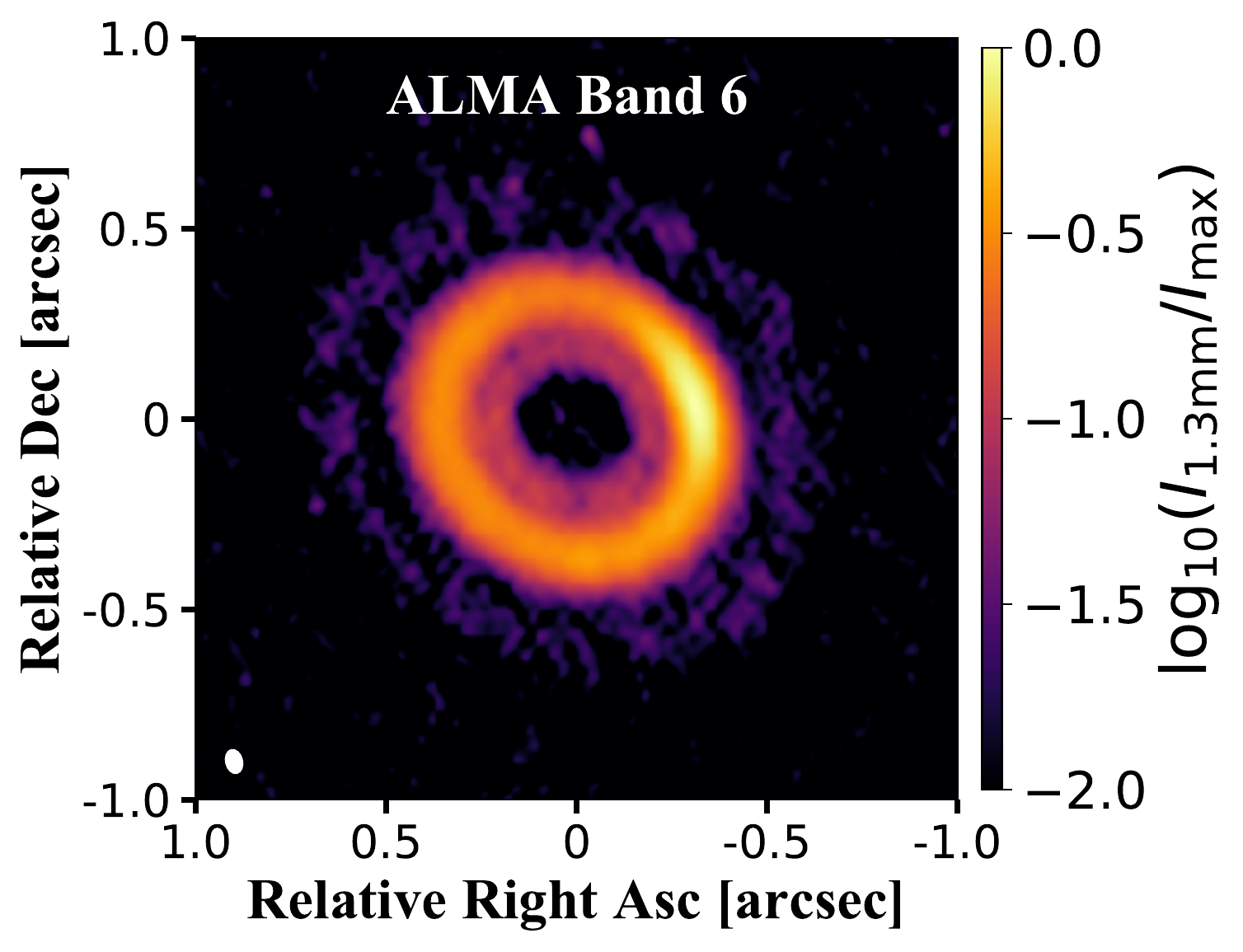}
    \caption{ALMA dust continuum observations in a color scale that highlights the faint structures.}
    \label{fig:ALMA_continuum2}
\end{figure}

The left panel of Fig.~\ref{fig:ALMA_data} shows the final image of the dust continuum emission at 1.3\,mm with a resolution of 0.06''$\times$0.04''. The same image is shown in  Fig.~\ref{fig:ALMA_continuum2} in a different stretch of the color scale that highlights the faint structures. The dust-continuum emission is mainly composed of a faint inner ring at $\sim0.19''$ ($\sim$60\,au), a bright and highly asymmetric ring (with a contrast of $\sim$4 by comparing the peak of the asymmetry with the opposite side) at $\sim0.35''$ (110\,au), and a much fainter ring at $\sim0.63''$ (200\,au; see also the radial profile of the continuum emission in Fig.~\ref{fig:ALMA_radial_profile}).  

From the image, the total flux that is enclosed in a circle with a 1.0''  radius from the center is 55.9\,mJy, with an uncertainty  of  7.7\,$\mu$Jy\,beam$^{-1}$. This flux is similar to the one obtained from the visibility fitting described later in this section of 56.1\,mJy. The azimuthally asymmetric structure encloses around 20\% of the total flux. By taking the flux within the contour of 40\% of the maximum (see top left panel in Fig.~\ref{fig:ALMA_SPHERE}), the flux within this structure is 12.3\,mJy.

\subsubsection{Optical depth and dust disk mass} 
We calculate the optical depth of the peak of the continuum ring, assuming \citep[e.g.,][]{dullemond2018}

\begin{equation} \label{eq:opticaldepth}
\begin{aligned}
    & \,I_\nu(r)=B_\nu(T_\mathrm{d}(r))(1-\exp{[-\tau_\nu(r)]}) \quad \mathrm{thus} \\
    &\, \tau = - \ln \left(1-\frac{I (r_\mathrm{peak})}{B(T_d(r_\mathrm{peak}))} \right),
\end{aligned}
\end{equation}

\noindent with $T_d$  being the dust temperature at the peak location ($r_{\rm{peak}}$). Equation~\ref{eq:opticaldepth} is only valid when neglecting dust scattering. In this scenario and using a dust temperature of  20\,K, we find that the optical depth at the peak (which is at the location of the asymmetry) is $\tau_{\rm{peak}, B6}=0.37$, similar to the values found in the DSHARP sample \citep[][]{huang2018, dullemond2018}. From the radiative transfer models in Sect.~\ref{sect:models_obs} that assume the results of the hydrodynamical simulations after 0.15\,Myr, the temperature at the midplane at the location of the asymmetry ($\sim110$\,au) is $36\,$K. Using this temperature for the calculation of the optical depth, we obtain $\tau_{\rm{peak}, B6}=0.16$. However, the emission may still be optically thick for two potential reasons. First, because the outer disk may be as cold as the interstellar medium ($\sim10$\,K), in which case $\tau_{\rm{peak}, B6}=1.8$. Second, dust scattering may not be negligible (which happens when dust grains have a radius comparable to the wavelength of the observations), in which case an optically thick region can be misidentified as optically thin \citep{zhu2019}.

Assuming that the emission is optically thin, we calculate the dust disk mass as $M_{\mathrm{dust}}\simeq\frac{{d^2 F_\nu}}{\kappa_\nu B_\nu (T)}$, where $d$ is the distance to the source, $F_\nu$ is the total flux at 1.3\,mm, and $B_\nu$ is the black-body surface brightness at a given temperature \citep{hildebrand1983}. Taking a mass absorption coefficient ($\kappa_\nu$) at a given frequency as $\kappa_\nu=2.3\,$cm$^{2}$\,g$^{-1}\times(\nu/230\,\rm{GHz})^{0.4}$ \citep{beckwith1990, andrews2013}, we obtain 168.5\,$M_\oplus$, and inside the asymmetry  the dust mass is  37\,$M_\oplus$ ($\sim$2.2\,$M_{\rm{Neptune}}$). Using the canonical dust-to-gas mass ratio of 0.01 \citep{mathis1977}, the disk mass is 0.05\,$M_\odot,$ and inside the asymmetry it is $\sim12$\,$M_{\rm{Jup}}$. However, assuming a dust-to-gas mass ratio of 0.01 inside the asymmetry may be unrealistic if this asymmetry is a vortex where particles are trapped. Based on the hydro-dynamical simulations of Sect.~\ref{sect:models_obs}, the dust-to-gas mass ratio at the location of the peak of the dust concentration is around 0.2, which leads to a mass of $\sim2.3$\,$M_{\rm{Jup}}$ inside the asymmetry.

\begin{figure}
    \centering
    \includegraphics[width=9.0cm]{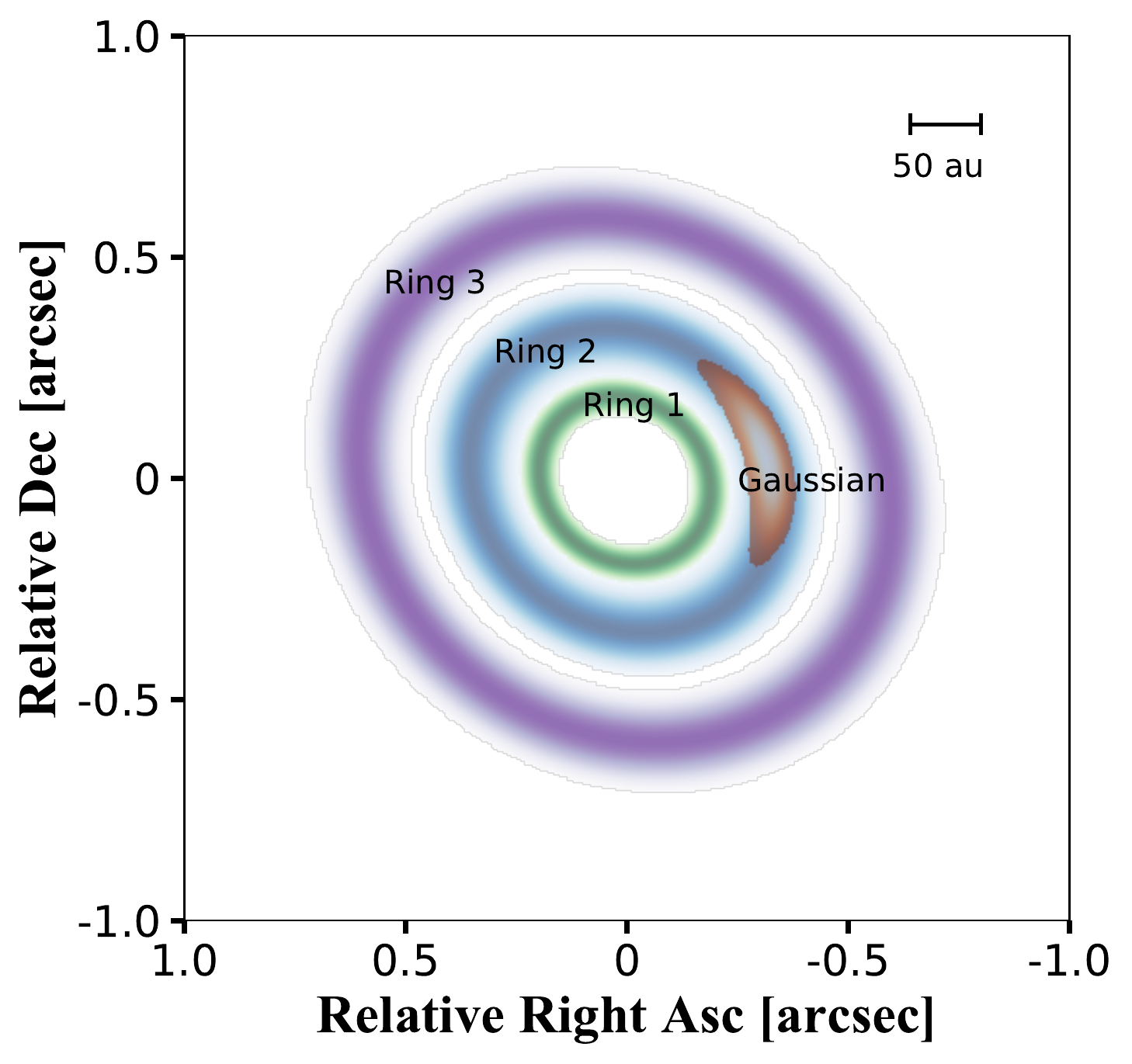}
    \caption{Schematic of components used for $uv$ fitting of the dust-continuum emission from ALMA.}
    \label{fig:schematic}
\end{figure}

\begin{table}
\centering
\caption{Best parameters from the $uv$-modeling. $R_{68}$ and $R_{90}$ denote the radius that encloses either 68\% or 90\% of the total flux ($F_\lambda$). Index $1$ corresponds to Ring 1, indeces $2$ and $4$ describe Ring 2, which is the main asymmetric ring, and index $3$ corresponds to Ring 3 (see
schematic in Fig~\ref{fig:schematic} for reference).   Pixel size is 5\,mas.}
\begin{tabular}{ c|c|c } 
  \hline
  \hline
\noalign{\smallskip}
    & Model     & units \\
\noalign{\smallskip}
  \hline
\noalign{\smallskip}
    $\delta_{\rm{RA}}$  & $-3.6_{-0.4}^{+0.4}$   & mas \\
    $\delta_{\rm{Dec}}$ & $-7.2_{-0.2}^{+0.4}$   & mas \\
    inc                 & $27.5_{-0.1}^{+0.3}$   & deg \\
    PA                  & $49.2_{-0.2}^{+0.8}$   & deg \\
\noalign{\smallskip}
  \hline
\noalign{\smallskip}
    $f_1$      & $1.01_{-0.05}^{+0.07}$   & (Jy/pix) \\
    $r_1$      & $202.7_{-1.4}^{+2.5}$    & mas \\
    $\sigma_1$ & $26.8_{-2.2}^{+1.7}$     & mas \\
\noalign{\smallskip}
    $f_2$         & $3.52_{-0.02}^{+0.03}$   & (Jy/pix) \\
    $r_2$         & $368.0_{-2.0}^{+2.2}$    & mas \\
    $\sigma_{2i}$ & $61.8_{-2.6}^{+2.2}$     & mas \\
    $\sigma_{2o}$ & $53.3_{-1.5}^{+1.6}$     & mas \\
\noalign{\smallskip}
    $f_3$      & $0.20_{-0.01}^{+0.02}$   & (Jy/pix) \\
    $r_3$      & $631.8_{-5.3}^{+1.9}$    & mas \\
    $\sigma_3$ & $62.3_{-6.9}^{+2.2}$     & mas \\
\noalign{\smallskip}
    $f_4$             & $8.29_{-0.15}^{+0.06}$   & (Jy/pix) \\
    $r_4$             & $349.8_{-0.1}^{+0.1}$    & mas \\
    $\theta_4$        & $-126.9_{-0.7}^{+0.4}$   & deg \\
    $\sigma_{4}$      & $31.5_{-0.1}^{+0.1}$     & mas \\
    $\sigma_{4\theta}$ & $21.8_{-0.2}^{+0.1}$     & deg \\
  \hline
\noalign{\smallskip}
    $R_{68}$      & $397.0 \pm 0.6$  & mas \\ 
    $R_{90}$      & $472.8 \pm 2.8$  & mas \\
    $F_\lambda$   & $56.1\pm0.2$  & mJy \\
\noalign{\smallskip}
  \hline
  \hline
\end{tabular}
\label{tab:mcmc_results}  
\end{table}

\subsubsection{Visibility fitting of the dust morphology}
We describe the dust-continuum emission observed with ALMA with a parametric model. Motivated by the radial profile from the \texttt{CLEAN} model image (see Fig.~\ref{fig:ALMA_radial_profile}), we describe \Lk~with three Gaussian rings and a Gaussian asymmetry in radius and azimuth direction as shown in Fig.~\ref{fig:schematic}.

For each model, the visibilities were obtained by optimizing the model profile with a spatial offset ($\delta_{\rm{RA}}$, $\delta_{\rm{Dec}}$), an inclination (inc), and position angle (PA), which are used to deproject the observational data. Therefore, each model has four extra free parameters in addition to those that describe the intensity profile. The Fourier transforms to obtain the model visibilities and the $\chi^2$ calculation are computed with the \texttt{galario} python package \citep{tazzari2018} using a pixel size of 5\,mas.

We sampled the posterior probability distribution with a Markov chain Monte Carlo (MCMC) routine based on the \texttt{emcee} python package \citep{emcee}. We use a flat prior probability distribution over a wide parameter range, such that the walkers would only be initially restricted by geometric considerations (inc $\in [0,90]$ , PA $\in [0,180]$, $\sigma \geq 0$).
We ran more than 250000 steps after convergence to find the most likely parameter set for each model, as well as taking the 16th and 84th percentiles for the error bars. Our results are shown in Table \ref{tab:mcmc_results}.

Based on the best parameters of this model, the faint inner ring is very narrow ($\sigma_1$=8.3$^{+0.5}_{-0.7}$\,au) andcentered at 63.6$^{+0.8}_{-0.4}$\,au. The width of this inner ring remains unresolved (the resolution of our observations in au is 19\,au$\times$12\,au). The main ring is described by two Gaussian profiles, one is a ring centered at 117.0$^{+0.7}_{-0.6}$\,au and is radially asymmetric, with the inner radial width slightly higher than the other width ($\sigma_{2i}$=19.4$^{+0.7}_{-0.8}$\,au vs. $\sigma_{2o}$=16.9$^{+0.5}_{-0.5}$\,au). The second Gaussian is an asymmetry that peaks at 111.2$^{+0.1}_{-0.1}$\,au, with an azimuthal width of 21.8$^{+0.1}_{-0.2}$deg. Finally, there is a faint ring at 200.9$^{+0.6}_{-1.7}$\,au with a width of 19.8$^{+0.7}_{-2.2}$\,au.

Figure~\ref{fig:ALMA_data} shows the comparison of the observations with the obtained $uv$-model, and the model after being imaged with the same procedure as the data. The right panel shows the residual image after subtracting the model from the observations; this is also imaged using \texttt{CLEAN}. The residuals map shows that the model describes the observations well, leaving residuals of a level of $10\,\sigma,$ mainly at the location of the asymmetry. It is interesting that the negative residuals line up with respect to the scattered light spiral arm in the south-west. However, this shape in the residuals may appear because we assume circular Gaussians in our model and the rings may be slightly eccentric. As a test, we also run a simulation with a Gaussian asymmetry that has different width in the azimuth direction \citep[e.g.,][]{perez2014, Cazzoletti2018}, but such a model does not significantly improve the residuals map. The model in Fig.~\ref{fig:ALMA_data} gives an azimuthal contrast of the asymmetry of $\sim4$. Figure~\ref{fig:ALMA_data_visi} shows the fit of this model of the binned data of the real and imaginary part of the deprojected visibilities.

\subsubsection{Emission of $^{12}$CO, $^{13}$CO, and C$^{18}$O} 
\begin{figure*} 
    \centering
    \includegraphics[width=18.0cm]{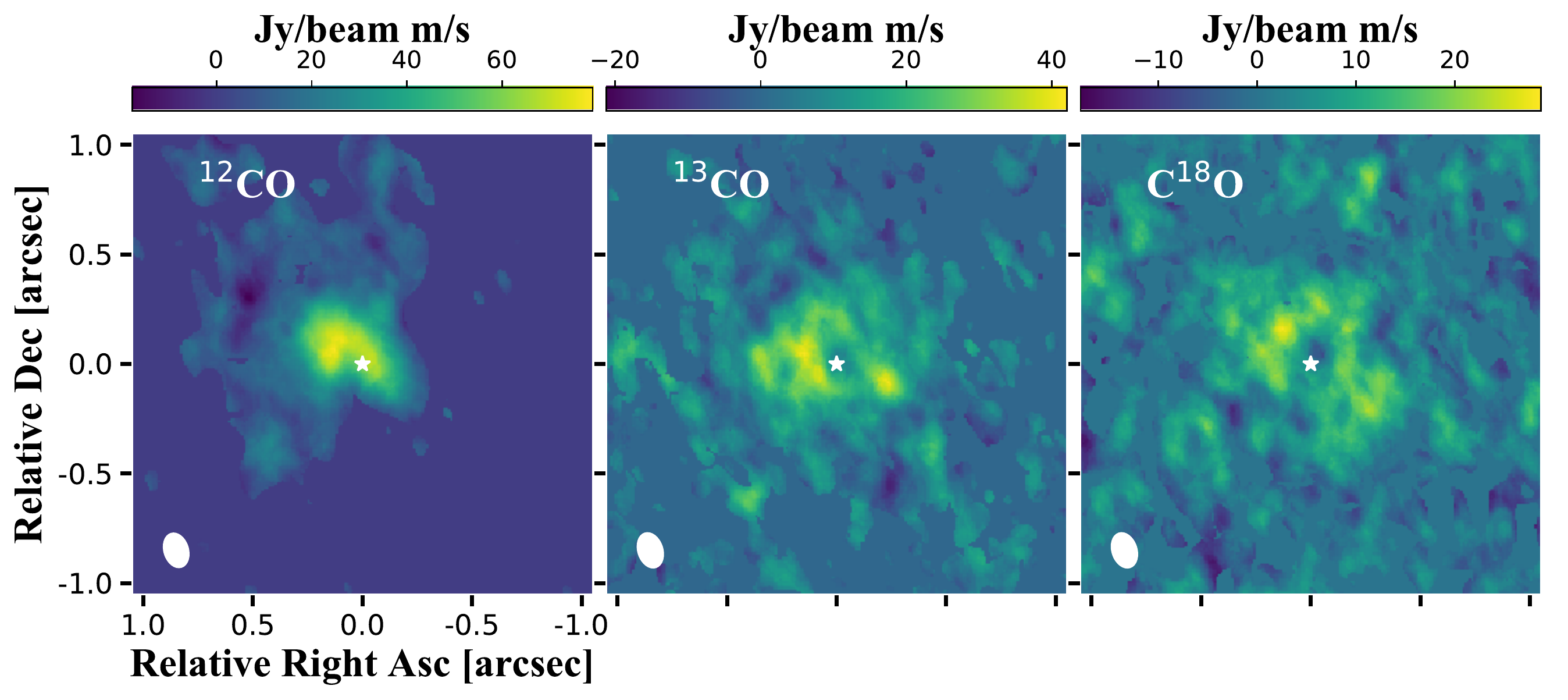}
    \caption{From left to right: moment 0 maps of the $^{12}$CO, $^{13}$CO, and C$^{18}$O, respectively, of \Lk~ from ALMA Band-6 observations.}
    \label{fig:ALMA_data_lines}
\end{figure*}

Figure~\ref{fig:ALMA_data_lines} shows the moment 0 maps of $^{12}$CO, $^{13}$CO, and C$^{18}$O. The emission from $^{12}$CO is mainly between the channel maps from 4 to 13.0\, km\,s$^{-1}$ (7 channels, see Fig.~\ref{fig:channel_maps}). The $^{12}$CO does not show an emission in the south-west that is as extended as it is in the north-east, and it is possible that this is because of  cloud contamination, and/or low signal-to-noise ratio due to the lack of short baseline observations that cover large scales. Thus, we cannot conclude that this asymmetry is real. 
For $^{13}$CO and $^{18}$CO, the emission mainly comes from five channels from 5.5 to 11.5\, km\,s$^{-1}$ (Fig.~\ref{fig:channel_maps}), and both look more azimuthally symmetric than the $^{12}$CO, although with a poor signal-to-noise ratio (for all three maps, the ratio between the peak and rms noise on the corresponding map is 5-6).

\begin{figure} 
    \centering
    \includegraphics[width=9.0cm]{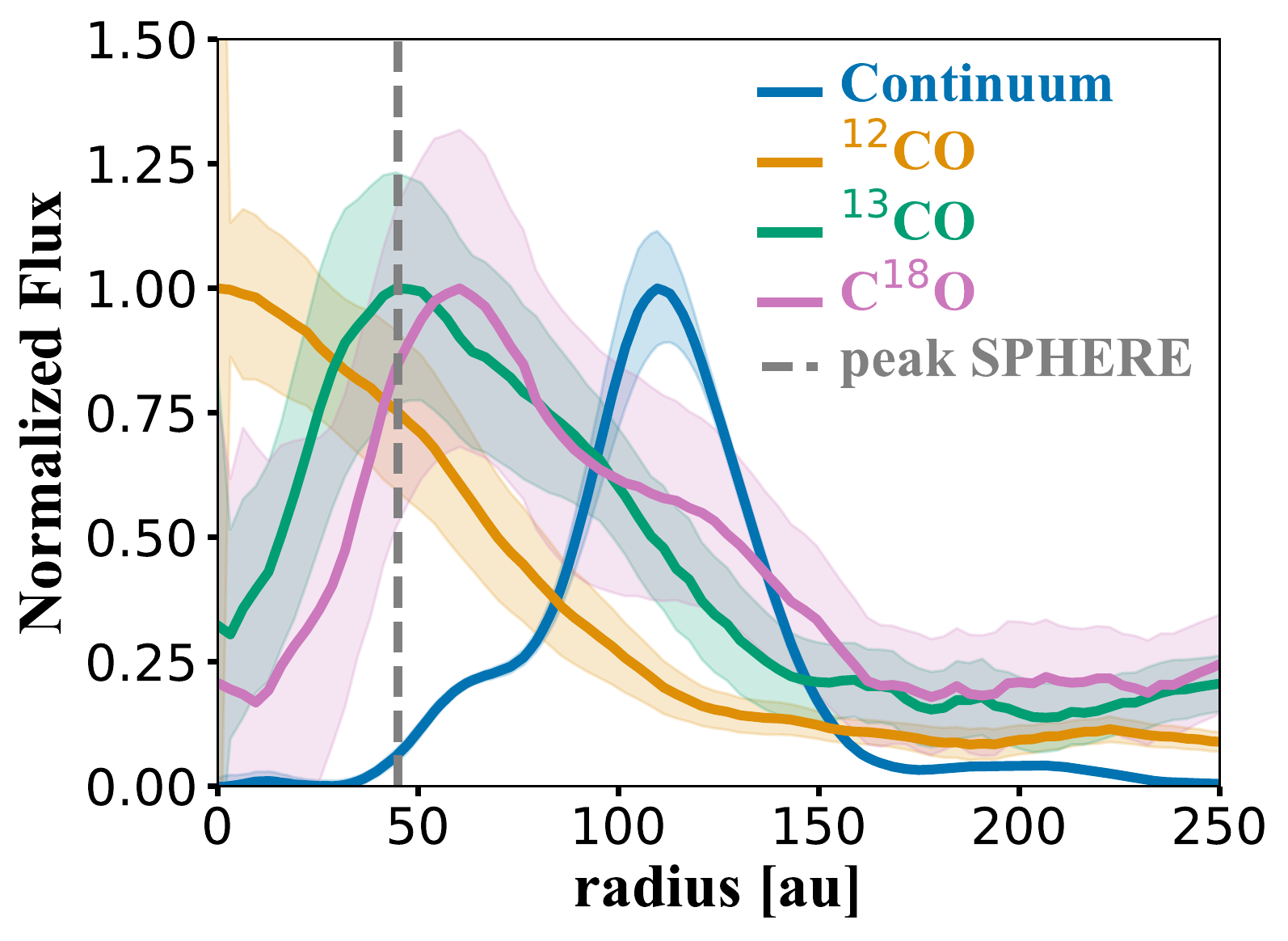}
    \caption{Azimuthally averaged radial intensity profiles of the deprojected  images of the continuum, and moment\,0 maps of the $^{12}$CO, $^{13}$CO, and C$^{18}$O of \Lk~ from ALMA Band-6 observations. Each profile is normalized to the peak. The shaded area is the standard deviation of each elliptical bin divided by the square root of the number of beams spanning the full azimuthal angle at each radial bin.}
    \label{fig:ALMA_radial_profile}
\end{figure}

Figure~\ref{fig:ALMA_radial_profile} shows the azimuthally averaged radial intensity profiles of the deprojected images of the continuum, and from the moment 0 maps of the $^{12}$CO, $^{13}$CO, and C$^{18}$O. Each profile is normalized to the peak. All 3 molecular lines peak inside the main peak of the continuum millimeter emission. In the moment 0 map of $^{12}$CO, it seems that there is an emission drop near the center, which could have been washed out by the noise and the azimuthally averaging in Fig~\ref{fig:ALMA_radial_profile}. The radial profile of the $^{12}$CO bends in the inner disk. Such a bending may be the combined effect of a reduced $^{12}$CO surface density and beam smearing \citep[e.g.,][]{bruderer2014,fedele2017,ubeira2019}. The radial profile of the  $^{13}$CO and C$^{18}$O show an inner drop of emission, where the $^{13}$CO peaks at a location similar to the inner ring observed with SPHERE ($\sim$45-50\,au), whereas the C$^{18}$O peaks at $\sim$60\,au,  which is very close to the peak of the inner faint ring observed with ALMA.  Figure~\ref{fig:moment8} shows the moment 8 map (peak value of the spectrum) of the $^{12}$CO, $^{13}$CO, and C$^{18}$O lines of \Lk. The moment~8 has been used to identify  gas substructures \citep[e.g.,][]{favre2019}, and in this case the three lines show a clear cavity in the moment~8 map.

\section{Comparison with planet-disk interaction models} \label{sect:models_obs}

\subsection{Estimation of planet mass and position} 
We investigate the origin of the cavity and the structures of \Lk~ in the context of planet-disk interaction. Based on the different radial extent of the cavity in scattered-light, CO molecular lines, and the dust continuum emission, it is possible to give an estimate of the mass of the potential planet carving this cavity, when assuming that the planet is in a circular orbit. As discussed in Sect.~\ref{sect:eccentricity}, planet eccentricity can affect different aspects of the disk, such as spiral shape, vortex survival, and gap size. 

There are two different approaches to obtaining such an estimation. First, by comparing the location of the cavity wall in scattered light versus the peak of the millimeter emission \citep[][]{ovelar2013}. The wall of the emission in scattered light is defined as the location where the intensity value is halfway between the intensity at the bottom of the gap and top of the ring. However, from our observations it is not possible to obtain the location of the minimum inside the cavity due to the coronagraph, and therefore we take the location of the peak (45\,au) as the wall of the cavity in scattered light. This provides us a lower limit of the planet mass. The peak of the millimeter emission is at 110;\, hence, $R_{\rm{wall}}/R_{\rm{mm, peak}}$ is 0.41, which will indicate a planetary mass higher than investigated in \cite{ovelar2013} (15\,$M_{\rm{Jup}}$ for a 1\,$M_\odot$ star), suggesting a brown dwarf-type companion. 

Another possibility is to take the gap size as observed from $^{13}$CO and compare it with the location of the peak from the continuum millimeter emission \citep{rosotti2016, facchini2018}. We note, however, that this approach is valid when the $^{13}$CO is optically thin. It is also difficult to obtain the location of the minimum flux inside the cavity from our $^{13}$CO due to the large uncertainties and poor resolution. Hence, we take the location where $^{13}$CO peaks ($\sim$50\,au) and provide a lower limit for the planet mass, using $(R_{\rm{mm}}-R_{^{13}\rm{CO}})/R_{^{13}\rm{CO}}=1.2$. Comparing with Fig.~11 from \cite{facchini2018}, this gives a planet-to-star mass ratio ($q$) between $4\times10^{-3}$ and $7\times10^{-3}$ (depending if the output from the simulations is taken at 1000 or 3000 planetary orbits, with a lower $q$ for longer times of evolution).

\begin{figure*} 
    \centering
    \includegraphics[width=15.0cm]{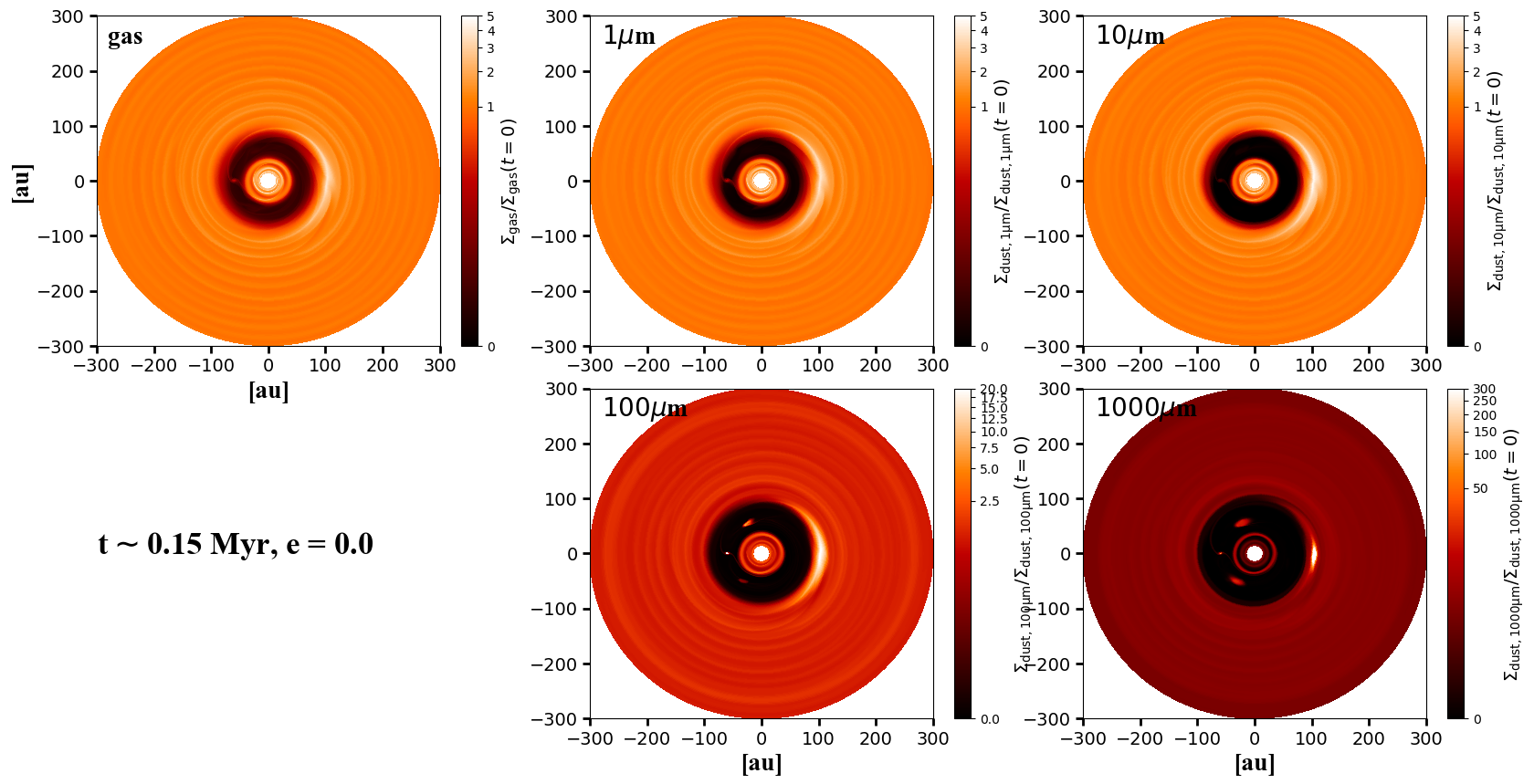}
    
    \includegraphics[width=15.0cm]{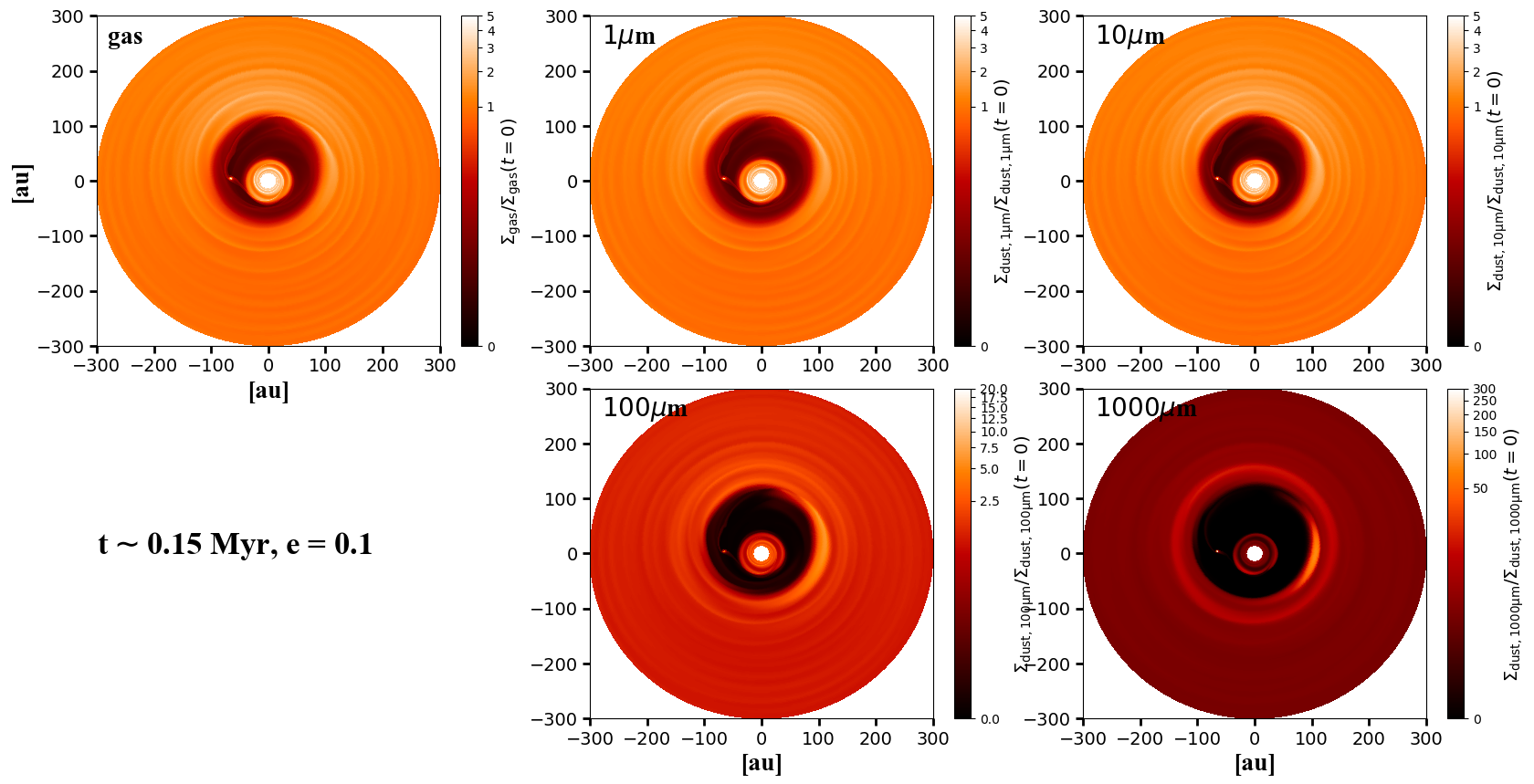}
    
    \includegraphics[width=15.0cm]{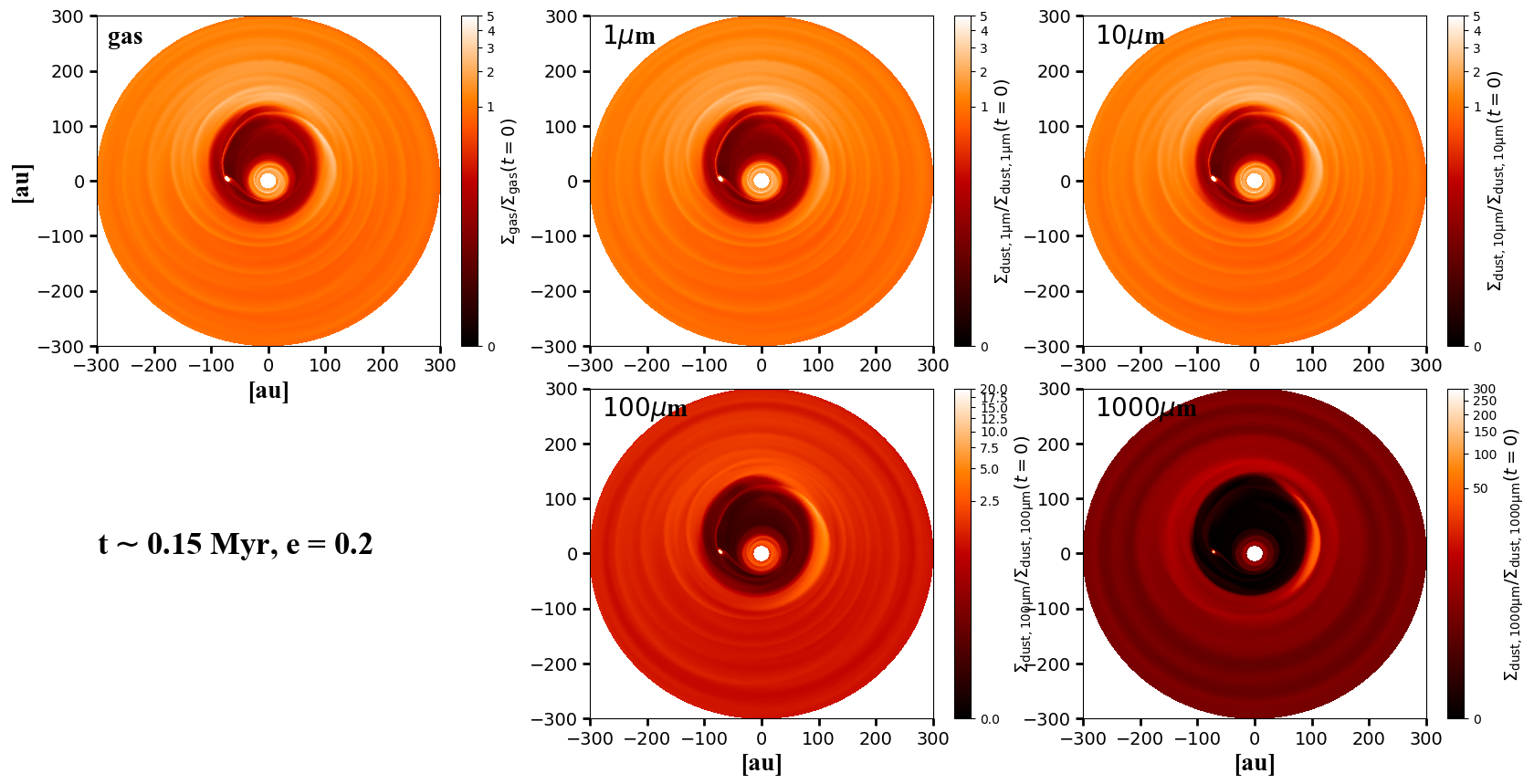}
    \caption{Results from hydrodynamical simulations performed with \texttt{FARGO3D} assuming a planet-to-star mass ratio of $4\times10^{-3}$ (10\,$M_{\rm{Jup}}$ around a 2.5\,$M_\odot$ star) at 60\,au and $\sim$0.15\,Myr of evolution ($\sim$500 orbits, where the exact output is taken when the vortex is opposite the planet). Each set of panels assumes a different planet eccentricity: $e=0.0, 0.1, 0.2$ from the top to the bottom. The simulation assumes an $\alpha$ viscosity of $10^{-4}$. For each eccentricity, the top left panel shows the gas surface density,  the panels from the top middle to the bottom right show the dust surface density of 1, 10, 100, and 1000\,$\mu$m-sized particles, respectively. }
    \label{fig:hydro_models}
\end{figure*}

Taking a planet-to-star mass ratio of $4\times10^{-3}$ and assuming the peak of the millimeter emission is radially located at around 7$R_{\rm{Hill}}$ from the planet position \citep[where $R_{\rm{Hill}}$ is the planet's Hill radius:][]{pinilla2012}, we obtain that the planet location is around 58 to 62\,au. We use 60\,au to perform hydrodynamical simulations of the gas and dust evolution and check if such a planet can create some of the observed structures: in particular, the large cavity and the asymmetric structure seen in the millimeter emission.  \cite{uyama2018} showed the Subaru/HiCIAO companion limits from a separation of 60\,au. The mass limits at 60\,au are well above (70-80\,$M_{\rm{Jup}}$) the values considered in our models. 

\subsection{Hydrodynamical simulations with \texttt{FARGO3D}}

We perform hydrodynamical simulations using the publicly available code \texttt{FARGO3D} \citep{benitez2016} and use the 2D version of the code (radial and azimuthal). We assume a locally isothermal disk and a power-law radial density profile: $\Sigma=\Sigma_0\,r^{-1}$. 

We use normalized units such that $G=M_{\star}+M_P=1$ and the location of the planet is at $r_p=1$. The simulations are performed from $r_{\mathrm{in}}=0.1$ to $r_{\mathrm{out}}=5.0$. We assume that the  planet's orbital semi-major axis is 60 au; thus, the radial grid spans from 6 to 300\,au, and it is  logarithmically spaced with 512 cells. The azimuth grid (from 0 to 2$\pi$) is linear with 1024 cells. The outer dust disk radius obtained from the visibility fit of the dust continuum emission is $\sim140$\,au, and most of the observed disks are two to three times larger in gas \citep{ansdell2018}, which supports our choice for the outer disk radius. 

The initial gas surface density $\Sigma_0$ at the position of the planet is such that the disk mass is $\sim$0.05\,$M_\odot$ (or 0.02\,$M_{\star}$, assuming the mass of the \Lk~star is 2.5\,$M_\odot$). This mass is consistent with the calculations from the dust-continuum emission (Sect.~\ref{sect:observations}). We assume a flared disk with a flaring index of 0.25 and a disk aspect ratio of 0.06 at the planet location (60\,au). This aspect ratio is obtained assuming that the temperature profile is \citep{kenyon1987}

\begin{equation}
        T(r)=T_\star\left(\frac{R_\star}{r}\right)^{1/2} \phi_{\rm{inc}}^{1/4},
  \label{eq:temp}
\end{equation}

\noindent where $R_\star=2\,R_\odot$ and $T_\star=5800$\,K as in \citet{andrews2011}. The incident angle $\phi_{\rm{inc}}$ is taken to be 0.05. The values of the aspect ratio and flaring index of our models agree with the best values of the models for fitting the spectral energy distribution (SED) of \Lk\, by \cite{andrews2011}. In these radiative transfer models, the scale height is 6.5\,au at 100\,au (i.e., an aspect ratio of 0.065), with a flaring index of 0.2.

The planet-to-star mass ratio is $4\times10^{-3}$ (10\,$M_{\rm{Jup}}$ around a 2.5\,$M_\odot$ star). We considere three values of planet eccentricity ($e$ = 0.0, 0.1, and 0.2). This choice is motivated by recent models that demonstrate that spiral arms launched by eccentric planets can change their pitch angle along the spiral \citep[e.g.,][]{zhu2022}, potentially appearing as two distinct spiral arms as in our SPHERE observations. Such planet mass is introduced in the first 100 orbits into the smooth disk. Planetary accretion and planet migration are not considered in these simulations. Depending on the planetary gas accretion rate, the gap shape (width and depth) can vary, and hence it can affect the potential formation of vortices at the edges of the carved gap when planet accretion  is considered \citep{bergez2020}. For massive planets such as those assumed in this work, 3D hydrodynamical simulations including planet migration have shown that a vortex can form at the outer edge of the planetary gap, diffusing material into the gap and migrating inwards with the planet \citep{lega2021}. Therefore, these two limitations can affect the interpretation of our models. The gravitational effect of the planet is smoothed out, such that the gravitational potential $\phi$  is softened over distances comparable to the disk scale height:

\begin{equation}
        \phi=-\frac{Gm_{p}}{(r^2+\epsilon^2)^{\frac{1}{2}}},
  \label{eq7}
\end{equation}

\noindent where $m_p$ is the planet mass and $\epsilon$ is taken to be $0.6h$ (where $h$ is the disk scale height defined as $h=c_s/\Omega$, with $c_s$ the sound speed and $\Omega$ the Keplerian frequency). 

Besides the gas evolution, we include the evolution of four dust species as in the \texttt{FARGO3D} version of \citet{benitez2019}, with the dust diffusion implementation from \citet{weber2019}. These grains have sizes of 1, 10, 100, and 1000\,$\mu$m. These particles are initially distributed as the gas with a dust-to-gas ratio of  0.01. We assume a power-law for the dust grain size distribution, such that $n(a)\propto a^{-3.5}$. The intrinsic volume density of the particles is assumed to be $\rho_s = 1.6~\mathrm{g\,cm}^{-3}$. Finally, we take an $\alpha$ viscosity \citep{shakura1973} of $10^{-4}$, in agreement with recent suggestions of low viscosity in disks \citep[e.g.,][]{flaherty2015, flaherty2017, teague2016}. This value of disk viscosity is also taken for the dust turbulent diffusion.  

\begin{figure*}
    \centering
        \includegraphics[width=18cm]{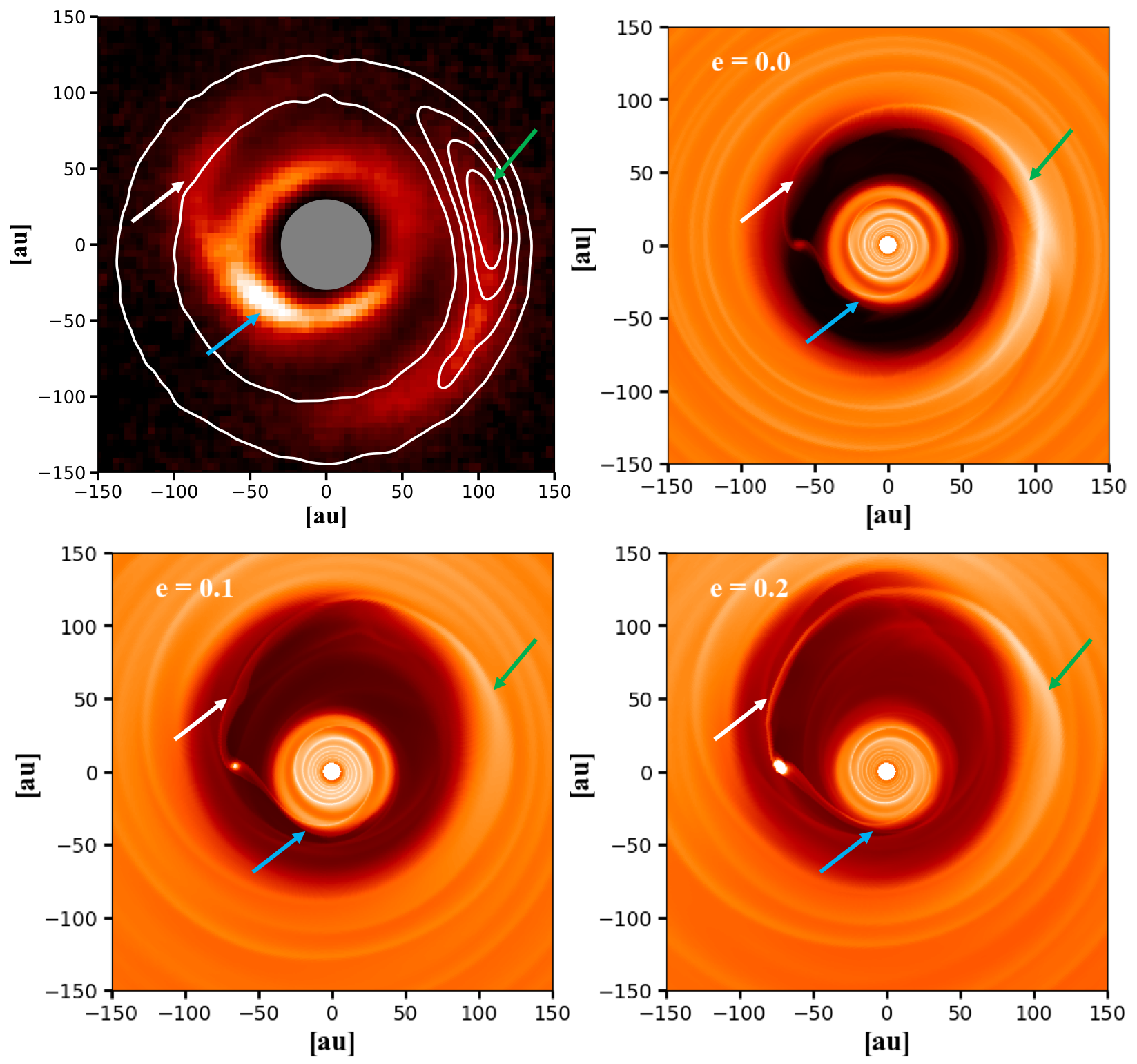}
    \caption{Comparison of models and observations. Upper left panel: Overlap of ALMA and SPHERE observations, both images are deprojected, and a distance of 318\,pc is used to change the units to \,au. The colors are SPHERE data of \Lk\, in the $H$-band ($Q_\phi \times r^2$) in linear scale, and the contours are ALMA data of \Lk\, in Band\,6 (1.3mm) every 20\%, 40\%, 60\%, and 80\% of the peak of the emission. Other panels: Zoomed-in view of the dust density distribution of 1\,$\mu$m-sized dust particles from the hydrodynamical simulations shown in Fig.~\ref{fig:hydro_models} for the three values of the eccentricity. The same coronograph as the SPHERE observations covers the inner disk. The arrows aim to qualitatively compare the observed structures (especially from SPHERE) with the models.}
    \label{fig:ALMA_SPHERE}
\end{figure*}

A summary of the results of our simulations is given in Fig.~\ref{fig:hydro_models}, which shows the gas surface density and the dust surface density for each grain size. All panels are normalized to the initial gas or dust surface density. We show the results for each eccentricity value after around 500\,orbits ($\sim$0.15\,Myr). The exact output that is selected for this figure is such that the vortex is opposite the planet location.  Fig.~\ref{fig:hydro_models2} shows the same as Fig.~\ref{fig:hydro_models}, but after 3000 orbits ($\sim$0.88\,Myr).

The planet triggers the RWI that leads to the formation of a vortex at the  outer edge of its gap at around 105-120\,au (the exact location depends on the planet eccentricity, being further away for higher planet eccentricity). This vortex appears in the gas and the small particles (1 and 10\,$\mu$m) with an azimuthal contrast of $3-3.5$. This contrast is much higher in the density of 100\,$\mu$m (contrast of 20) and 1000\,$\mu$m (contrast $>$ 600) dust particles due to particle trapping in the vortex \citep[e.g.,][]{ataiee2013}.

For the case of the planet with zero eccentricity, the vortex starts to dissipate after the first 700 orbits because of  the disk’s turbulent viscosity, and after around 1000 orbits there is no signature of the vortex either in the gas surface density or in the density of the small particles (1 and 10\,$\mu$m). Nonetheless, the concentration of the large particles (100  and 1000\,$\mu$m) inside the vortex takes much longer to decay, and after 3000 orbits (see Fig.~\ref{fig:hydro_models2}), this concentration remains. Interestingly, at early times of evolution, besides the asymmetry at $\sim$105\,au, there is also an outer ring at $\sim$130\,au. At later times (Fig.~\ref{fig:hydro_models2}), these signatures are more evident in the dust density map of the 1mm dust grains, with an asymmetric ring at around 120\,au and an additional ring outside around 150\,au. 

\begin{figure*}
    \centering
      \includegraphics[width=18cm]{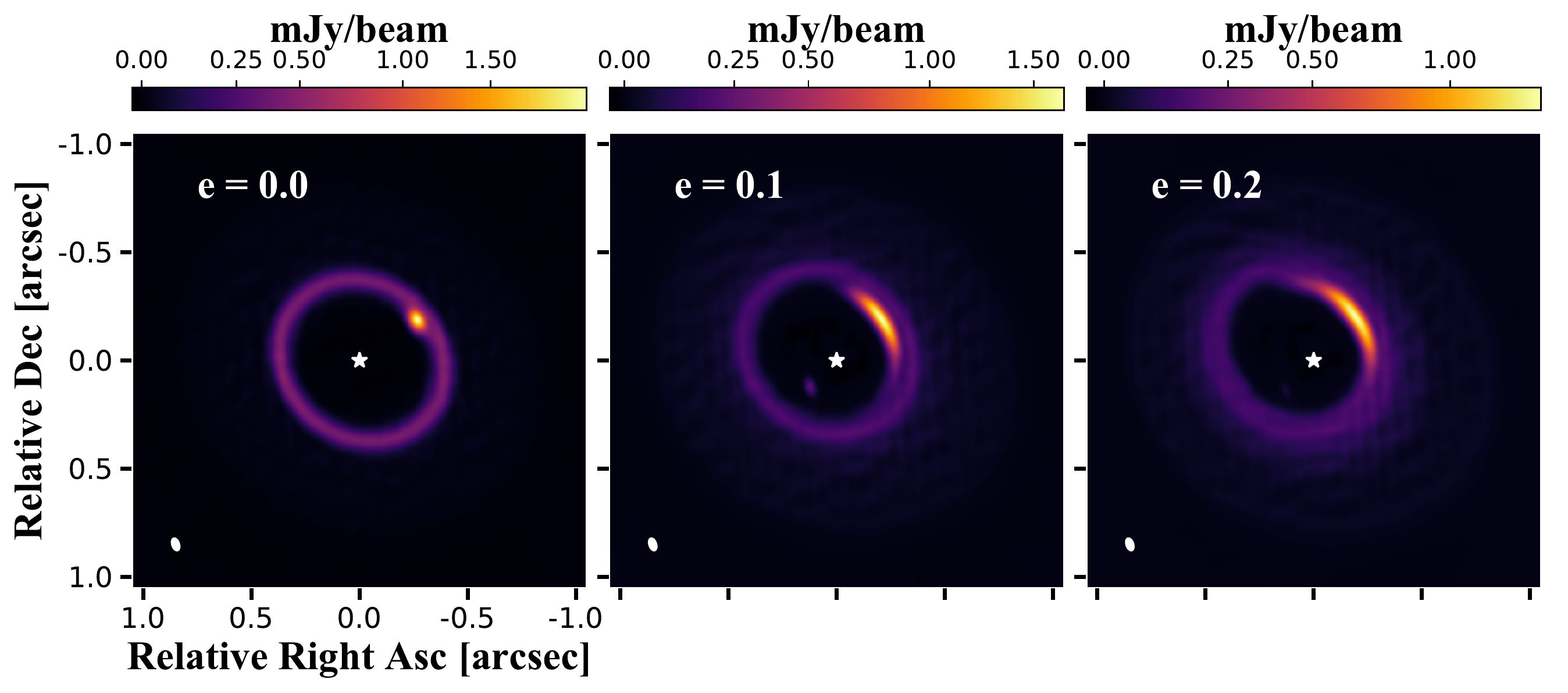}
    \caption{Synthetic image at 1.3\,mm from combining the results from FARGO with radiative transfer calculations and after creating the synthetic images with \texttt{SIMIO}.}
    \label{fig:ALMA_synthetic_image}
\end{figure*}

For the case where the planet eccentricity is 0.1, the vortex lives for $\sim$500 orbits in the gas and hence in the distribution of the small-sized particles (1 and 10\,$\mu$m); but, as in the case of zero eccentricity, the concentration of the large sized particles takes longer to dissipate. In the case of $e=0.1$, the  asymmetry in the large particles (100  and 1000\,$\mu$m) lives until around 700 orbits, and then the concentration becomes a ring-like structure that merges with the outer ring, as is shown in Fig.~\ref{fig:hydro_models2}, where a clear ring-like structure is formed in the density
of the 100 and 1000\,$\mu$m-sized particles at $\sim140$\,au.

For the case of $e=0.2$ a similar situation occurs. The vortex in the gas density and the concentration of the small-sized particles (1 and 10\,$\mu$m) survives until around $\sim400$ orbits, and in the large grains it remains until $\sim600$ orbits. The ring-like structure that remains is initially very eccentric, but it circularizes with time, as seen when comparing Fig.~\ref{fig:hydro_models} and Fig.~\ref{fig:hydro_models2}. Based on these results, we hypothesize that if the planet is in an eccentric orbit, it must be very young to explain the azimuthal asymmetry observed with ALMA.

Planets are a natural explanation for the formation of spiral arms and the structures observed in near-infrared scattered light \citep[e.g.,][]{bae_zhu2018}. However, it has been  shown that to reproduce the contrast of these spiral arms as observed, 3D simulations are required because 2D simulations usually underestimate their brightness \citep{juhasz2015,dong2017}.

Because of  the limitations in comparing our 2D simulations with the observations in scattered light, we only performed a visual inspection of the spiral arms launched by the proposed planet and qualitatively compared it with the scattered-light observations. Fig.~\ref{fig:ALMA_SPHERE} shows the overlap of ALMA and SPHERE observations. Both images are deprojected, and we use a distance to the source of 318\,pc to show the scale in au. We compare the observations with the results from the hydrodynamical simulations and show a zoomed-in view of the dust density distribution of 1\,$\mu$m-sized dust particles for the three values of the eccentricity.

The planet launches three spiral arms, two inner spiral arms, and one outer spiral arm, in agreement with the results from \cite{bae_zhu2018b}. In the models where the planet has some eccentricity, the spiral arms are distorted, in particular the outer spiral arm, which shows that the pitch angle suddenly changes values in different locations, as shown by \cite{zhu2022}. In this comparison, part of the inner ring observed with SPHERE is part of the inner spiral, which could explain the non-uniform brightness distribution of this ring. The outer spiral could be the one observed with SPHERE in the north-east, which passes through the vortex, making it very prominent again in the (south-) west. As we explain in Sect.~\ref{sect:observations}, the pitch angle of the two spiral arms is similar, which may indicate that the origin is the same and the difference may originate from the distortion expected when the planet is in an eccentric orbit and/or when it passes through the vortex. The launching points of our two spiral arms inferred from the SPHERE observations are further away from the planet position ($\sim$90\,au), and a possible explanation is that the inner ring at $\sim$45\,au is hot and puffed up, blocking the starlight near the planet.

\subsection{Radiative transfer and comparison with the dust-continuum emission from ALMA} \label{sect:radmc}

In order to compare the results from the hydrodynamical simulations to  the ALMA observations, we perform radiative transfer calculations with \texttt{RADMC3D} \citep{dullemond2012}. We calculate the opacity of each grain size from the \texttt{FARGO} simulations considering the DSHARP opacities \citep{birnstiel2018} and using \texttt{optool} \citep{dominik2021}. We assume a black-body radiation field from the central star as the radiation source and use $1\times10^{7}$ photons and $5\times10^{6}$ scattering photons for our calculations.

To calculate the total volume dust density, we follow

\begin{equation}
        \rho_{\mathrm{d}}(R,\varphi,z, \mathrm{St}) = \frac{\Sigma_{\mathrm{d}}(R, \mathrm{St})}{\sqrt{2\,\pi}\,h_{\mathrm{d}} (R, \mathrm{St})}\,\exp \left( -\frac{z^2}{2\,h_{\mathrm{d}}^2(R, \mathrm{St})} \right)\,,
        \label{eq:volume_density}
\end{equation}

\noindent where  $z = r\,\cos(\theta)$ and $R=r\,\sin(\theta)$, with $\theta$ being a polar angle. We keeep the same radial and azimuthal resolutions as for the hydrodynamical simulations. For the vertical grid, we use 128 cells. The particle scale height $h_{\mathrm{d}}$ is given by \citep{youdin2007, birnstiel2010}

\begin{equation}
        h_{\mathrm{d}}(\mathrm{St})=h \times \rm{min} \left( 1,\sqrt{\frac{\alpha}{\mathrm{min}(\rm{St},1/2)(1+\rm{St}^2)}}\,\right),
        \label{eq:dust_scaleheight}
\end{equation}

\noindent where \rm{St} is the Stokes number of the dust particles calculated at the midplane, that is $\rm{St}=\frac{a\rho_s}{\Sigma_g}\frac{\pi}{2}$. Under the assumptions of our model, a 1mm particle at $\sim$110au has St$\sim$0.1, implying that the scale height of the 1mm grains is around 3\% of the disk scale height. Thus, millimeter-sized particles are well confined in the midplane, whereas the micron-sized particles have St$\sim10^{-4}$ in the outer region, and therefore their scale height is almost the same as the gas.

We obtaine the temperature profile for each grain size and calculated  images at 1.3\,mm. We assumed the distance, PA, and disk inclination of \Lk.  To create realistic ALMA images with the same $uv$ coverage as the actual observations, we use the \texttt{SIMIO} package \footnote{\url{https://www.nicolaskurtovic.com/simio/}}, which replaces the observed visibilities with the radiative transfer model visibilities. Before the radiative transfer modelling, we remove the emission from the inner disk ($<$50\,au), which is not detected with ALMA. In simulations that include the growth and fragmentation of dust particles, the inner disk is expected to be depleted of dust in around one million years, when the gap carved by the planet efficiently filtered dust particles from the outer disk, while the dust initially located inwards of the planet grows and efficiently drifts toward the star \citep[e.g.,][]{pinilla2016}. 

Figure~\ref{fig:ALMA_synthetic_image} 
shows the comparison between the model and observations when taking the outputs from the hydrodynamical simulations after 500 orbits (Fig.~\ref{fig:hydro_models}), for different planet eccentricities. For $e=0.0$, the dust is highly concentrated in the center of the vortex, creating a more compact asymmetry compared to observations; this is the case even after 3000 orbits (Fig.~\ref{fig:hydro_models2}). For the case of $e=0.2$, the asymmetry is similar to that observed with ALMA; however, the cavity is very eccentric in contrast to the ALMA observations. In the case of $e=0.1$, the main asymmetry is surrounded by a ring, and after convolution with the beam, the two (the asymmetry and the ring) almost merge. A point-like structure is obtained at the location of the planet, which remains in the simulations due to the lack of proper planet accretion \citep{bergez2020}. 

\section{Discussion} \label{sect:discussion}

\subsection{Different radial distribution of the scattered light, millimeter emission, and CO lines} \label{sect:segregation}

The disk around \Lk~shows a large segregation (radial and azimuthal) of the distribution of small grains traced with scattered light, the large grains traced with the dust-continuum emission from ALMA, and the gas distribution potentially traced with the emission from CO isotopologues. The radial difference in these distributions can be seen in Fig.~\ref{fig:ALMA_radial_profile}. In the top left panel of Fig.~\ref{fig:ALMA_SPHERE}, we overlap the ALMA and SPHERE observations to highlight the different structures at the two wavelengths, in particular in the azimuthal direction, where the location of the end of the south-west spiral in scattered light coincides with the location of the asymmetry observed with ALMA. The potential connection between these structures is discussed in Sect.~\ref{origin_spirals}.

The radial difference in the distribution of gas and small/large dust particles is typical in observations of transition disks \citep[e.g.,][]{dong2012, marel2016, villenave2019}, and it is expected from planet-disk interaction models as shown in Sect.~\ref{sect:models_obs}. In the assumption of our models, the planet is located at 60\,au in order to have the asymmetric ring at a similar location to the observations ($\sim$110\,au). This simulation succeeds in explaining the asymmetry and potentially the formation of a faint outer ring as observed, although at a different location, $\sim$140\,au in the models versus 200\,au in the observations. Therefore, after convolution with the ALMA beam, the asymmetry and the outer ring obtained in the models almost merge. 

This model is roughly consistent with the distribution of the small dust particles and the gas. In the observations, $^{12}$CO, which is usually optically thick, blends in the inner disk, while the $^{13}$CO and the scattered light peak at the same location at around 45-50\,au (Fig.~\ref{fig:ALMA_radial_profile}). The fact that these two peaks coincide supports the idea that $^{13}$CO is also optically thick and traces variations in disk temperature (as the scattered light) instead of gas surface-density variations. The hydrodynamical models presented in Sect.~\ref{sect:models_obs} do predict a faint ring (in gas and in small dust species) inside the planet gap (see zoomed-in images in the panels in Fig.~\ref{fig:ALMA_SPHERE}). The inner edge of the gap is located at 45\,au. This ring is not fully symmetric due to the launched spiral arms inside the planet's orbit, and it could potentially explain the ring and its brightness variations observed in scattered light. However, when we perform radiative transfer models and create images at 1.6\,$\mu$m to compare with observations, the ring from the synthetic images looks mostly symmetric (Fig.~\ref{fig:syntheticH}), but this is likely due to the fact that our simulations are 2D and cannot produce the high contrast of the spirals obtained from more realistic 3D simulations. Similarly, in the 1.6\,$\mu$m synthetic images the outer spiral arm is also very faint, and the vortex dominates the emission (Fig.~\ref{fig:syntheticH}). A possible way to mitigate this inconsistency is to use another equation of state. We use vertically isothermal disks, while with an adiabatic equation and long cooling times the contrast of the spirals is expected to increase, making possible to detect them even with ALMA observations with high levels of sensitivity and high resolutions \citep{Speedie2022}.

As we mention in Sect.~\ref{sect:radmc}, in simulations where dust growth is also included, it is expected that this inner disk is not long-lived because particles grow and quickly drift inwards. One possible solution is that fragmentation of particles is efficient in these regions, making it possible to continuously keep small  (micron-sized) particles in the inner disk that are well coupled to the gas, which remain invisible at millimeter emission. 

The $^{12}$CO, $^{13}$CO, and C$^{18}$O lines seem to be less abundant in the inner disk. These molecular line observations have a very low signal-to-noise-ratio, and they lack short baselines, so the nature of these emission lines in the inner disk is poorly constrained from observations.  The $^{13}$CO and C$^{18}$O peak close to the location where the planet is assumed in the models (60\,au). In our simulations, there is very little material in the co-rotation region of the planet, and it is insignificant compared to the material inside and outside the planet's orbit. In fact, this co-rotation material is only expected to be observable for low-mass planets that do not open a deep gap \citep[e.g.,][]{perez2019, weber2019}, creating at least three observable rings. A possible explanation is that the peak of  $^{13}$CO and C$^{18}$O is tracing the location where the gas surface density starts to increase (instead of the actual peak of the gas density), possibly because both of these lines may not be fully optically thin. In the simulations, the gas surface density starts to increase at around 65\,au reaching its maximum at the location of the vortex. Nonetheless, to test this idea, thermochemical simulations are needed, which are not included in our models. Higher sensitivity observations are needed to better constrain the shape of the CO emission and its isotopologues and see if their emission agrees with the existence of a real gas cavity or an actual gap in the gas surface density. 

Planets are not the only possible explanation for the radial segregation seen between the gas and small/large dust particles. The presence of a dead zone interplaying with a magneto-hydrodynamical wind \citep{pinilla2016_DZ} or a photoevaporative wind \citep{garate2021} can also explain such differences. In addition, at the outer edge of a dead zone, vortices can be formed due to the RWI as well \citep{flock2015}. Furthermore, variations of the disk viscosity can also trigger spiral arms \citep{lyra2015}, but in this case numerous spiral arms would be expected, while only two are detected in current observations. However, the extension of a dead zone for a star such as \Lk~ is expected to be around 20\,au \citep{delage2021}, which is much smaller than the observed cavity. A clear way to distinguish this scenario from the planet scenario is to actually detect potential planet(s) or their circumplanetary disks inside the cavity, as in the case of PDS\,70 \citep{keppler2018, benisty2021}. 

\subsection{Origin of the spiral arms and millimeter asymmetry} \label{origin_spirals}

The pitch angles of observed spiral arms are larger compared to the ones obtained from models of planet-disk interaction, in particular when comparing with the spirals expected outside the planet's orbit. The pitch angle of the spiral arms is directly connected to the local scale height of the disk, that is the disk temperature. Typically, to obtain the large observed pitch angles, the disk temperatures need to be unrealistically high \citep[][]{benisty2015}.  In our hydrodynamical simulations we have a similar problem and the spiral arms look tighter in the simulations when compared to the SPHERE observations. In addition, the disk temperature is vertically stratified such that the surface is hotter than the midplane. Spirals in 3D simulations adopting vertical temperature stratification have larger pitch angles in the surface than the midplane \citep[e.g.,][]{juhasz2018}.

One way to reconcile this discrepancy is to assume that the planet is located sufficiently far outside the observed spirals, and the primary and secondary arms inside the planet's orbit are the ones that we observe \citep[as suggested in the case of MWC\,758, HD\,135344B and HD\,100453][]{dong2015, fung2015}.  Applying this idea to \Lk, a possible solution is that a planet is located between the rings observed with ALMA at 110 and 200\,au, creating the spiral arms inside its orbit, while the asymmetry may be explained by a vortex at the inner edge of this hypothetical planet, and where dust particles are accumulated in a faint, ring-shaped emission at the outer edge of this gap. Such a scenario cannot explain the formation of the cavity itself (and the observed radial segregation of the gas and dust particles), as we aim to do with the hydrodynamical simulations presented in this paper. Therefore, a potential scenario is the combination of two planets, as investigated by \cite{Baruteau2019} for the case of MWC\,758.

Another possibility for the formation of the spiral arms is that the disk is massive and cold enough to be gravitationally unstable, forming several spiral arms in the disk \citep{lodato2004}. Such spiral arms could be observed in scattered light, as well in the millimeter emission \citep[e.g.,][]{dipierro2015}, and the observed pitch angles for \Lk\, can also be explained by gravitational instability \citep[][]{baehr2021}. The disk mass obtained from the dust continuum emission is such that the disk-to-stellar mass ratio is 0.02, and the Toomre parameter \citep{toomre1964} is well above unity in the disk (assuming the temperature profile of Eq.\ref{eq:temp}). However, the calculation of the disk mass is highly uncertain when using the millimeter flux, due to the assumptions of the optical depth, dust opacities, dust temperature, and dust-to-gas mass ratio.  A potential diagnostic to determine if \Lk~ may be gravitationally unstable is to detect signatures in the disk kinematics, where, unlike in the planet-disk interaction case, the gravitational-instability spirals perturb the velocity in the entire disk, creating wiggles that are visible at all disk radii and all azimuthal angles \citep{hall2020}.

For the origin of the asymmetry, we explored the possibility of a vortex triggered by the RWI in the context of planet-disk interaction. As mentioned in the previous section, another possibility is a vortex formed by the same instability at the edge of a dead zone. Besides these possibilities, disk eccentricity can create azimuthal overdensinty features, as in the cases where there is a binary companion \citep[e.g.][]{calcino2019, ragusa2020}. For \Lk, there are no observational signatures of a binary companion \citep{uyama2018}. 

Finally, the vortex itself may trigger the spiral arms \citep{Paardekooper2010, chametla2022}. From the millimeter fluxes, we inferred that the mass inside the asymmetry could be a few Jupiter masses (Sect.~\ref{sect:observations}). This is a tentative alternative because of the overlap of the south-west spiral arm with the location of the asymmetry. However, the spirals triggered by a vortex are expected to be weak density waves and impossible to detect in scattered-light observations \citep{huang2019}.

\subsection{Origin of the faint rings around the main asymmetric structure}

Recent high angular resolution observations of transition disks have unveiled that the ring-shaped accumulation of dust particles around the cavity of these disks is a composite of more complex substructures. Some examples are the cases of the transition disks around LkCa\,15, 2MASS J16100501-2132318 \citep{facchini2020}, HD\,135344B \citep{Cazzoletti2018}, SR\,21 \citep{muro2020}, and PDS\,70 \citep{keppler2019, benisty2021}. In addition, in some of these observations a very faint ring has also been detected in the outer regions far away from the main structures, as in the cases of HD\,100546 \citep{walsh2014, fedele2021, max2021}, AA\,Tau \citep{loomis2017}, HD\,97048 \citep{gerrit2017}, and DM\,Tau \citep{kudo2018}.

\cite{facchini2020} suggested that the equation of state considered in the simulations can lead to the production of one or multiple rings around the main cavity. In a vertically isothermal disk, the interaction with a planet can lead to multiple rings, while in an adiabatic disk with large cooling factors and the same disk and planet parameters can lead to a single ring. In that paper, where the authors present ALMA observations of LkCa\,15 and 2MASS J16100501-2132318, the main ring is composed of two rings, where the inner ring is brighter than the outer ring, which contradicts what we see in our observations of \Lk; hence, it is unclear if adiabatic simulations of planet-disk interaction can produce the right brightness distribution for \Lk.

Another degeneracy concerning the number of rings that a single planet can create comes from the assumed  planet mass. For a low-mass planet, the material in the co-rotation region can be observable, creating at least three rings  \citep[one inside, one at the co-rotation region, and one outside the planet; e.g.,][]{bae2017, dong2017b, perez2019}. As we discuss in Sect.~\ref{sect:segregation}, a low-mass planet cannot explain the formation of the cavity itself, but it is still possible that there is a low-mass planet between 60 and 110\,au and/or between 110  and 200\,au.

An additional degeneracy about the number of rings is the inclusion of dust growth and fragmentation \citep[e.g.,][]{bae2018, bergez2022}. In this case in particular, the  dust turbulent parameter plays a key role because if dust is highly diffused in a disk, even when a pressure trap is present, dust particles may not accumulate in the pressure bump \citep{ovelar2016}. In addition, faint rings around the main accumulation of the large grains can be formed due to the ring's self-evolution effect on the disk’s thermal structure \citep{zhang2021}.

Finally, for the formation of a very faint ring in the outer disk, a possible scenario is that there is an outer planet trapping the particles, but such a planet must form late in the evolution, such that most of the dust has drifted inwards and little dust is outside to be trapped \citep{pinilla2015}. Alternatively, if the initial disk gas surface-density distribution is a power law tapered with an exponential function, then any planet that is located outside the cutoff radius  can trap only the little amount of dust that is outside \citep{max2021}.  In our models, we do reproduce an outer ring without the need for an extra planet, but this is closer in than observed. 

\subsection{Effect of planet eccentricity} \label{sect:eccentricity}

Planet eccentricity adds another degeneracy to planet-disk interaction models as investigated by \cite{chen2021}, where a lower mass planet in an eccentric orbit can create a similar gap shape as a more massive planet. Planet eccentricity can affect the vortex by smoothing the outer gap edge, breaking the RWI condition for its formation. \cite{DAngelo2006} and \cite{Hosseinbor2007} investigated how the disk-planet interaction can affect the gap and planet eccentricities. They found that an eccentric planet carves a shallower and broader gap, implying that the density profile at the gap edge is less steep than the ones of a gap opened by a planet on a circular orbit. Therefore, it is harder to obtain the RWI condition in a disk with an eccentric planet.  \cite{Hosseinbor2007} proposed that an eccentric planet is not able to affect the disk morphology if $e<R_{H}/r_{p}=(q/3)^{1/3}$ (where $R_H$ is the Hill's radius of the planet, $r_p$ is the planet location, and $q$ is the planet-to-star mass ratio). This critical value for our planet mass is 0.11. As our results show, an eccentricity of 0.1 already has an effect on the vortex's survival and the dust concentration. In our simulations, the vortex lives over shorter timescales, and therefore the dust concentration is more azimuthally extended and has lower contrast once the vortex starts to dissipate. In addition, the planet eccentricity can also affect the shape of the launched spiral arms as we discuss earlier in the paper, whereby the spiral pitch angle can change along the spiral. 

Finally, \cite{duffell2015} showed that the depth of the gaseous gap created by a Jovian planet can be reduced by one order of magnitude when the planet is in an eccentric orbit with values of $e=0.1$. This can have direct consequences on the emission of CO and isotopolgues observed inside the cavity, but thermochemical models coupled with hydrodynamical simulations are required  to properly quantify this effect.

\section{Conclusions} \label{sect:conclusions}
In this paper, we present new scattered light SPHERE observations at $J-$ Band (1.2\,$\mu$m) and $H-$ Band (1.6\,$\mu$m), in addition to new ALMA observations in Band\,6 (1.3\,mm) of the transition disk around \Lk. These observations are compared to hydrodynamical simulations that include gas and dust evolution with the goal of explaining the observed structures with a single planet, which does neither migrates nor accretes material. The main conclusions are the following.

\begin{itemize}
    \item The SPHERE observations reveal two types of clear structures. First, a non-uniform ring in brightness at around 45\,au from the star, with brightness variations along the ring of $\sim$50\%. Second, two spiral arms, one in the north-east and the other in the south-west with similar pitch angles ($\sim$9-11$^{\circ}$) and radial launching points ($\sim$90\,au). However, there is a high uncertainty on the pitch angles of the spirals in particular of the south-west due to the unknown geometry of the scattered light (the inclination and position angle is assumed from the dust continuum emission). 
    \item The ALMA observations of the dust-continuum emission reveal three main structures: a large cavity surrounded by a faint inner ring at around 60\,au; a bright asymmetric ring at around 110\,au, and this asymmetry has an azimuthal width of around 20$^\circ$; in addition to a faint ring at around 200\,au.
    \item The $^{12}$CO, $^{13}$CO, and C$^{18}$O lines seem to be less abundant in the inner disk. All of  these lines peak inside the main ring observed in the dust-continuum emission (at 110\,au). The $^{13}$CO peaks at a similar location to the inner ring observed with SPHERE ($\sim$45\,au), while the  C$^{18}$O peaks around 60\,au, which is very close to the faint inner ring observed in the dust continuum with ALMA. Any conclusions about the gas distribution from these CO observations must be taken with caution because they have poor signal-to-noise ratios due to the lack of short baseline observations and short integration times.
    \item The radial segregation in the distribution of the gas and small and large dust particles can be reproduced when assuming a 10\,$M_{\rm{Jup}}$ planet located at 60\,au. Such planet mass is well below the current observational limits for planetary companions at such distances \citep{uyama2018}.
    \item Our qualitative comparison of the observations with hydrodynamical simulations suggests that to explain the asymmetry seen with ALMA, the planet should be in an eccentric orbit with $e=0.1$. A planet in a circular orbit leads to a very narrow azimuthal concentration of the particles compared to observations, whereas a planet in a more eccentric orbit leads to a very eccentric cavity during the timescales when the asymmetry is still present. At longer times, the cavity circularizes, but the asymmetry also decays. The results from these models suggest that the planet is young.
    \item According to our comparison with hydrodynamical simulations, it is possible that the two spiral arms observed with SPHERE originate from the outer spiral launched by the proposed eccentric planet, which corresponds to the spiral in the north-east. When the spiral passes through the vortex, it becomes very prominent again in the (south-) west. In this scenario of the eccentric planet, the pitch angle changes along the spiral in addition to the distortion when it overlaps with the vortex, explaining why in the observations one single spiral may appear as two. 
\end{itemize}

Our results show that \Lk\, is an exciting target to search for (eccentric) planets while they are still embedded in their parental disk, making it an excellent candidate for planet-disk interaction studies.

%%%%%%%%%%%%
\section*{Acknowledgements}
We are thankful to the referee for the constructive report.
P.P. and N.T.K. acknowledges support provided by the Alexander von Humboldt Foundation in the framework of the Sofja Kovalevskaja Award endowed by the Federal Ministry of Education and Research. P.P. acknowledges the Cluster of Excellence STRUCTURES for providing a baby-office during the developing of this paper.  This project has received funding from the European Research Council (ERC) under the European Union’s Horizon 2020 research and innovation programme (PROTOPLANETS, grant agreement No. 101002188). Support for J. H. was provided by NASA through the NASA Hubble Fellowship grant \#HST-HF2-51460.001-A awarded by the Space Telescope Science Institute, which is operated by the Association of Universities for Research in Astronomy, Inc., for NASA, under contract NAS5-26555. G.R. acknowledges support from an STFC Ernest Rutherford Fellowship (grant number ST/T003855/1). M.V. research was supported by an appointment to the NASA Postdoctoral Program at the NASA Jet Propulsion Laboratory, administered by Oak Ridge Associated Universities under contract with NASA.
The authors are thankful with the developers of \texttt{galario}, \texttt{emcee}, \texttt{FARGO3D}, \texttt{RADMC3D}, and \texttt{optool} for making their codes publicly available. This paper makes use of the following ALMA data: ADS/JAO.ALMA\#2018.1.00302.S.  ALMA is a partnership of ESO (representing its member states), NSF (USA) and NINS (Japan), together with NRC (Canada), MOST and ASIAA (Taiwan), and KASI (Republic of Korea), in cooperation with the Republic of Chile. The Joint ALMA Observatory is operated by ESO, AUI/NRAO and NAOJ.
  
%%%%%%%%%%%%%%

\bibliographystyle{aa} % style aa.bst
\bibliography{ref}

\begin{appendix}
\section{Visibility fit}
\begin{figure}[h!]
    \centering
    \includegraphics[width=9.0cm]{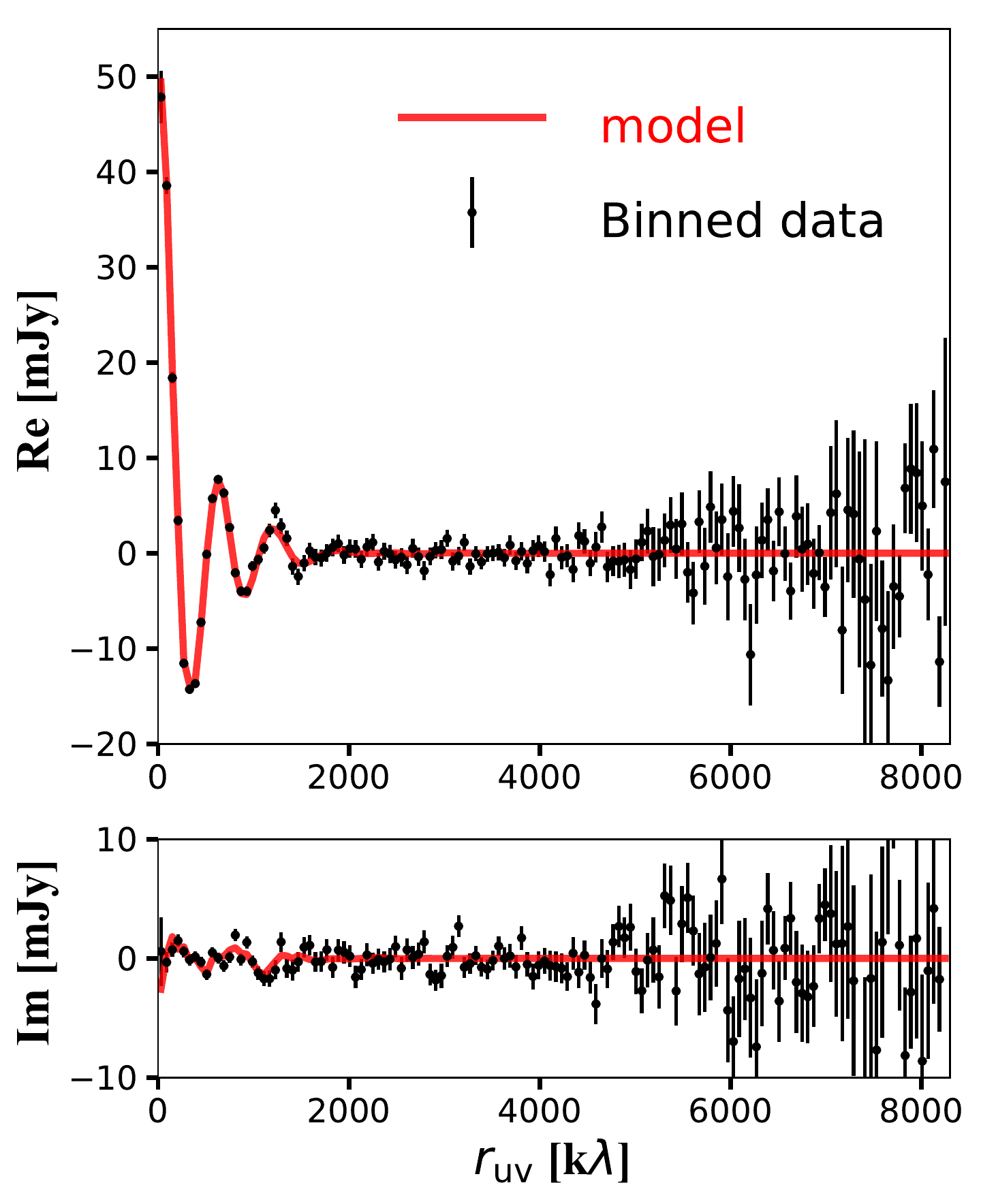}
    \caption{Real (upper panel) and imaginary (lower panel) part of the binned and deprojected visibilities versus the model with the best-fitting parameters from \texttt{galario} (red solid line). The error bars correspond to the standard error in each bin.}
    \label{fig:ALMA_data_visi}
\end{figure}

Figure~\ref{fig:ALMA_data_visi} shows the fit to the real and imaginary part of the visibilities to the model with the best-fitting parameters from \texttt{galario}. 
\FloatBarrier

\section{Channel maps and moment 8 maps} \label{Appendix:maps}
\begin{figure*}
   \centering
    \tabcolsep=0.05cm 
    \begin{tabular}{cc}   
        \includegraphics[width=9cm]{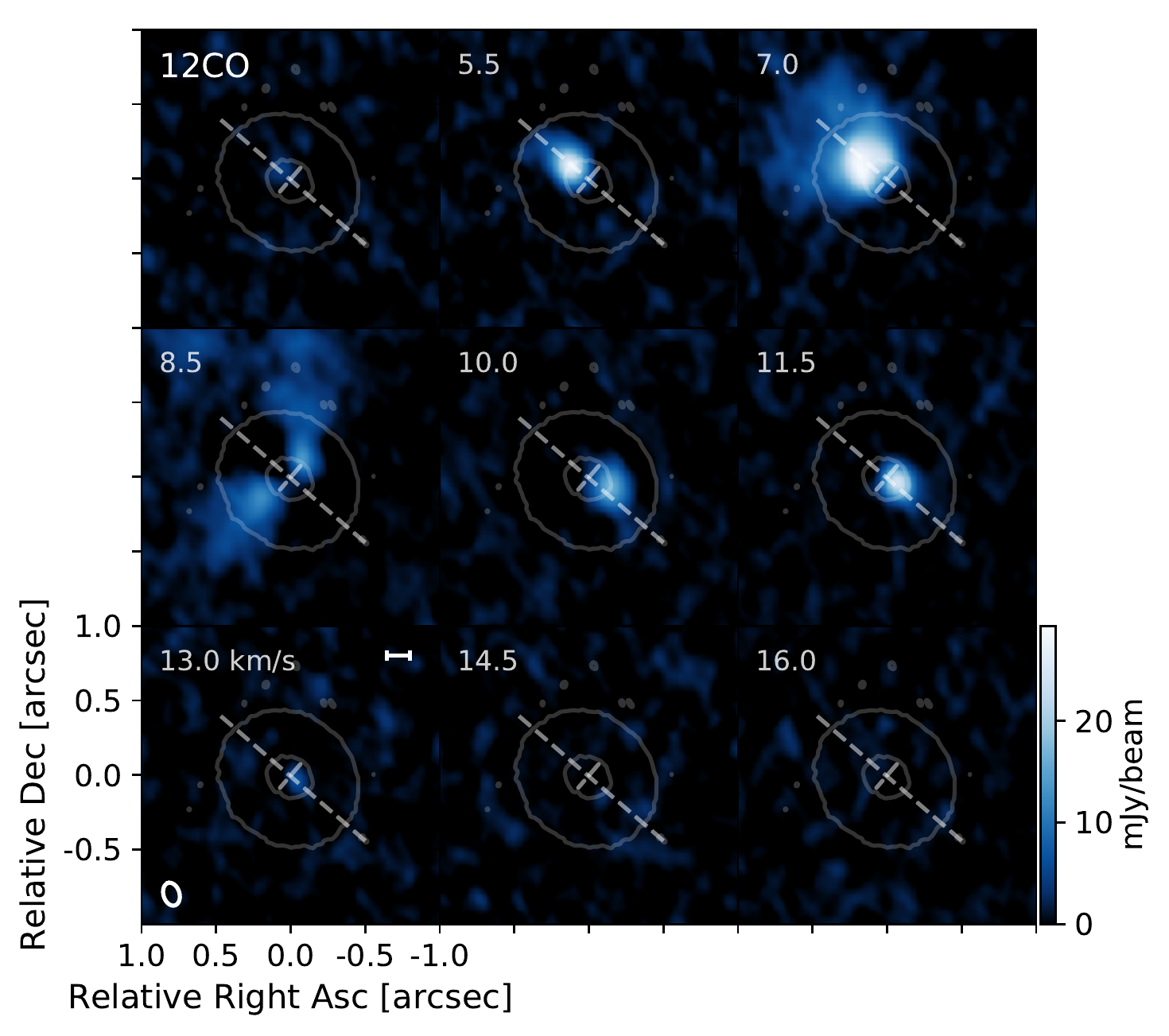}&
        \includegraphics[width=9cm]{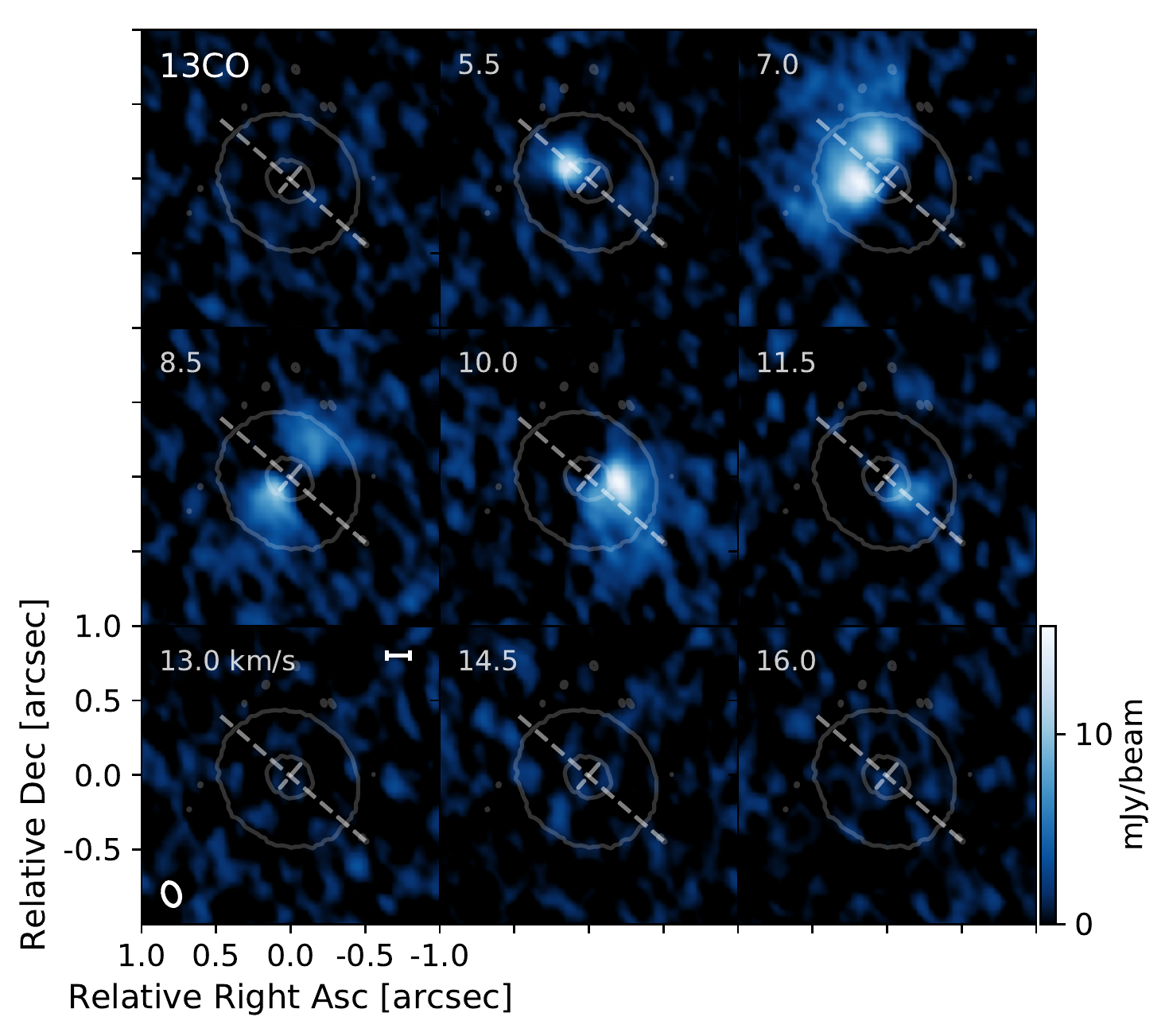}\\
        \includegraphics[width=9cm]{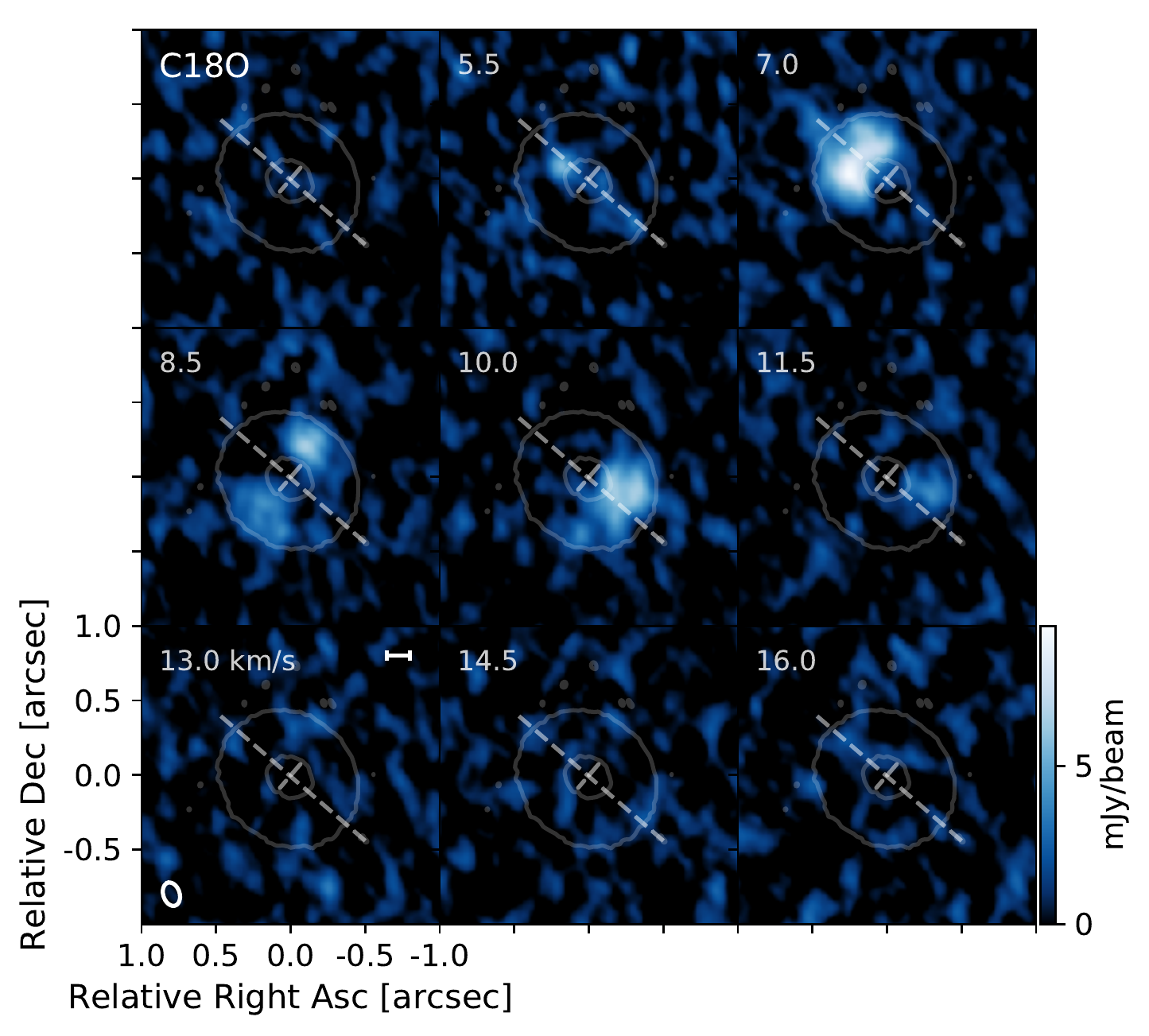}&
    \end{tabular}
    \caption{Channel maps of $^{12}$CO, $^{13}$CO, and C$^{18}$O of \Lk. The contours are at the $5\times\sigma$ level of the continuum emission. The scale bar in the left panel represents a scale of 50\,au.}
    \label{fig:channel_maps}
\end{figure*}

\begin{figure*}
   \centering
        \includegraphics[width=18cm]{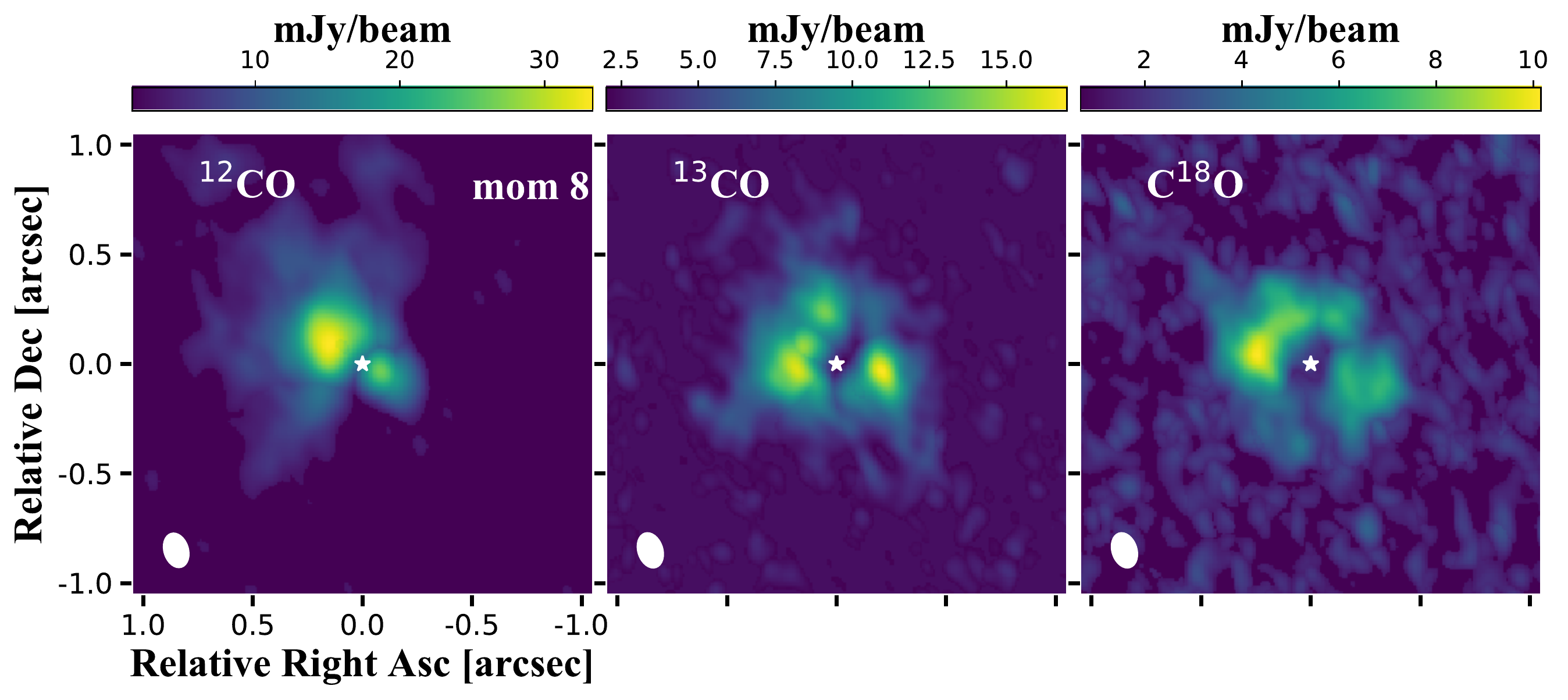}
    \caption{Moment 8 map (peak value of the spectrum) of the $^{12}$CO, $^{13}$CO, and C$^{18}$O lines of \Lk.}
    \label{fig:moment8}
\end{figure*}

Figure~\ref{fig:channel_maps} shows the channel maps of the $^{12}$CO, $^{13}$CO, and C$^{18}$O of \Lk\, from our ALMA observations. Figure~\ref{fig:moment8} shows the moment 8 maps (peak value of the spectrum) of the $^{12}$CO, $^{13}$CO, and C$^{18}$O lines of \Lk.

\FloatBarrier

\section{Hydrodynamical simulations at longer times of evolution}

\begin{figure*} 
    \centering
    \includegraphics[width=15.0cm]{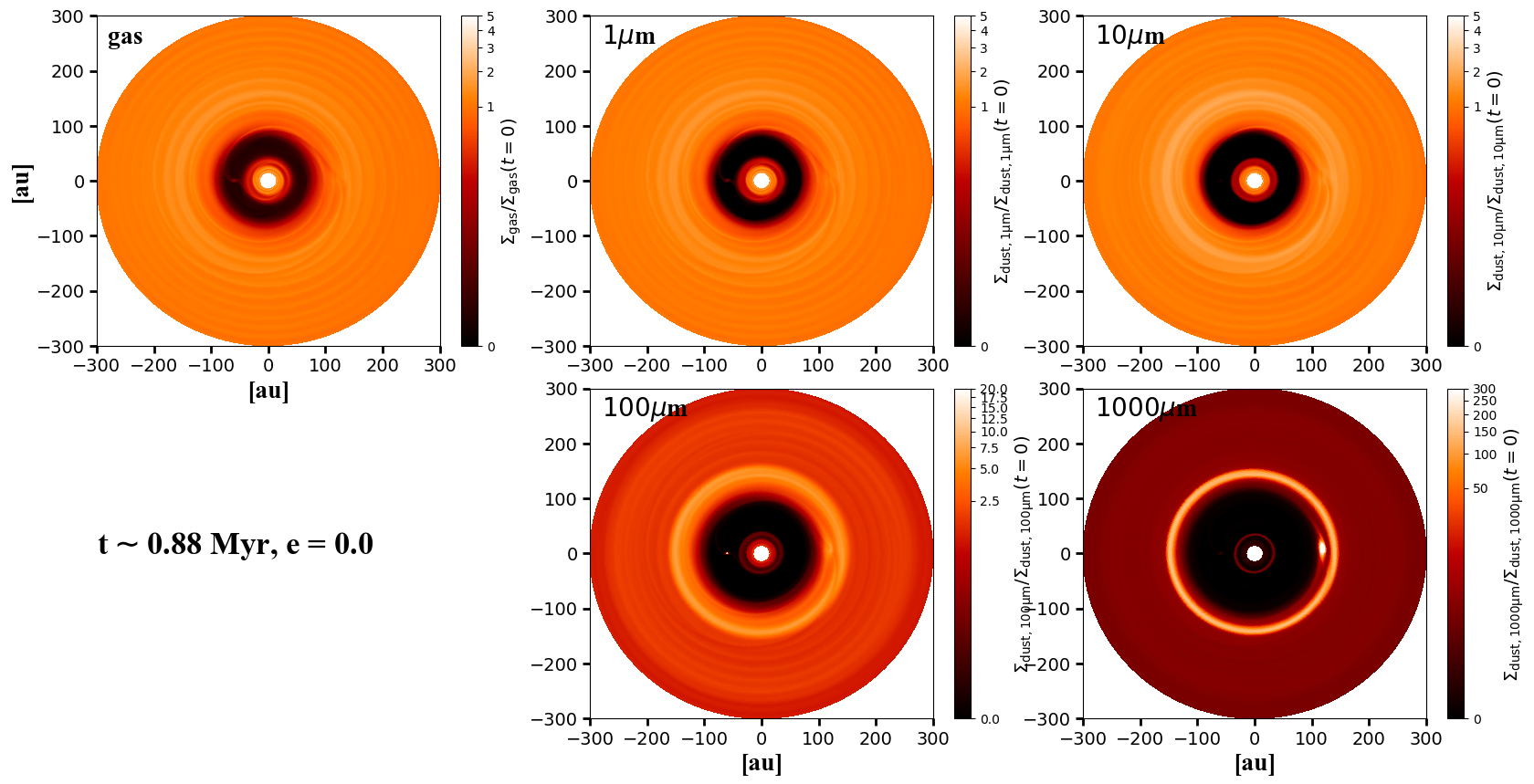}
    
    \includegraphics[width=15.0cm]{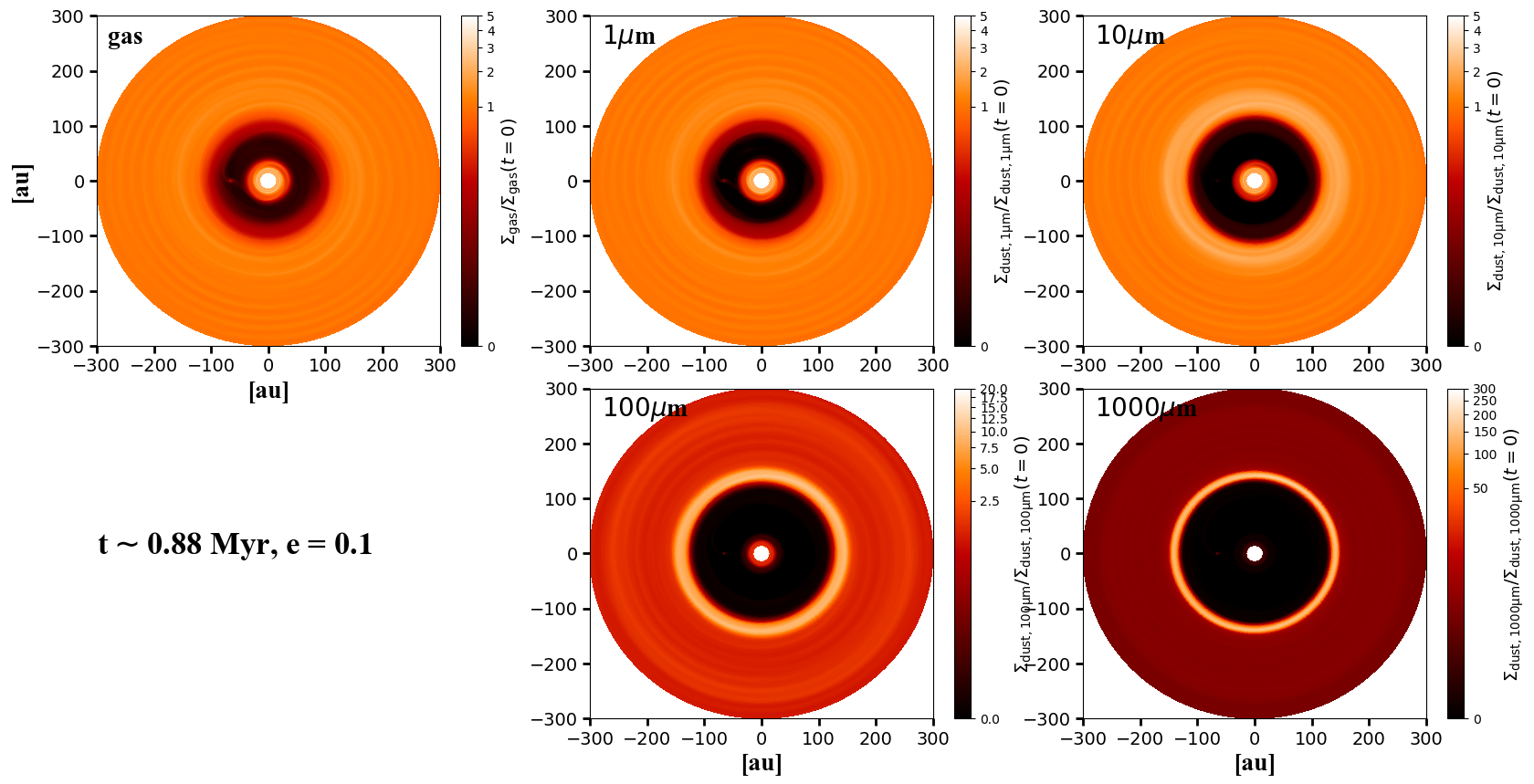}
    
    \includegraphics[width=15.0cm]{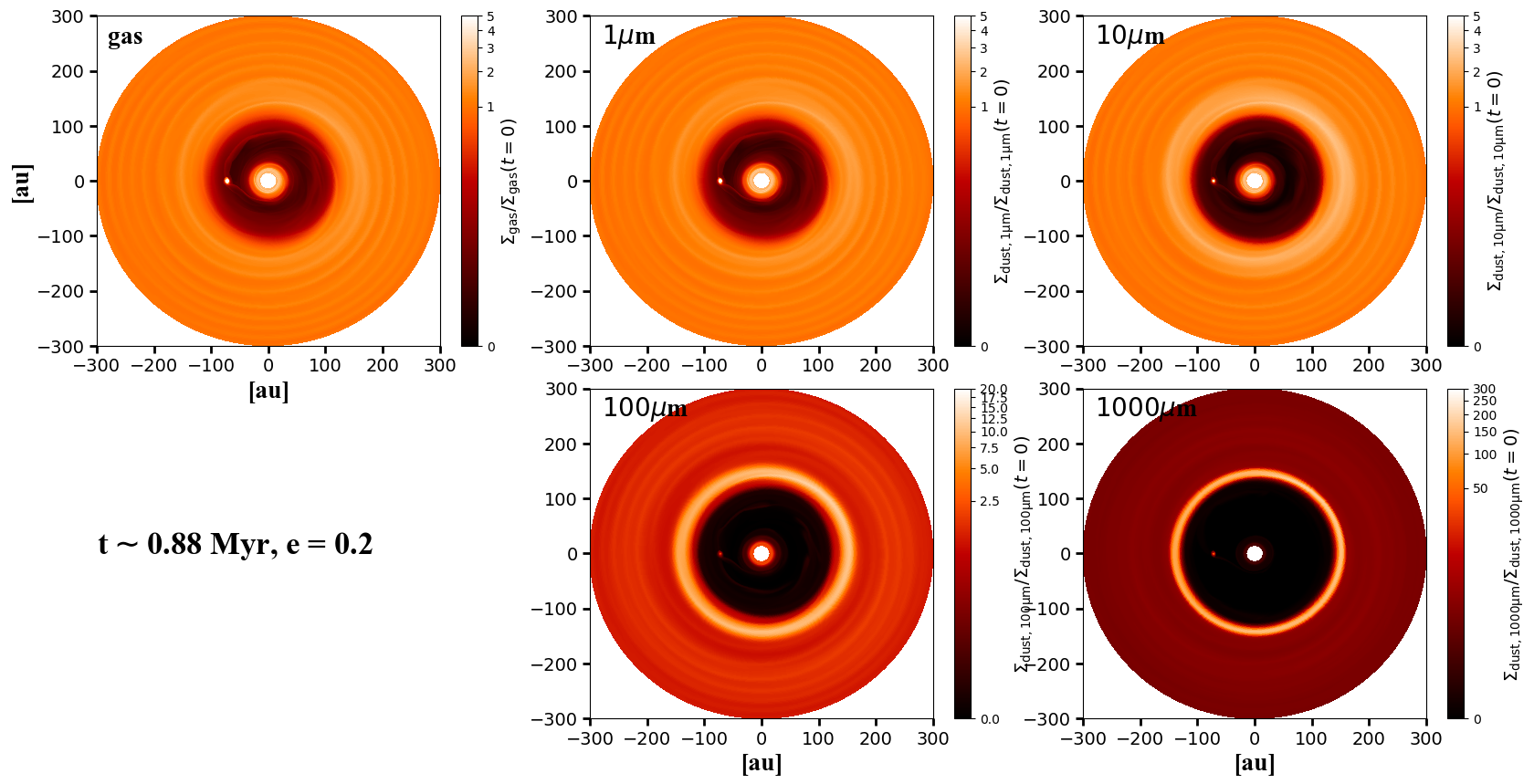}
    \caption{Results from hydrodynamical simulations as in Fig.~\ref{fig:hydro_models},  but at 0.88\,Myr of evolution ($\sim$3000 orbits).}
    \label{fig:hydro_models2}
\end{figure*}

Figure~\ref{fig:hydro_models2} shows the results from hydrodynamical simulations as in Fig.~\ref{fig:hydro_models}, but at 0.88\,Myr of evolution ($\sim$3000 orbits).

\FloatBarrier

\section{Synthetic image in H-band}
\begin{figure}[h!]
    \centering
    \includegraphics[width=9.0cm]{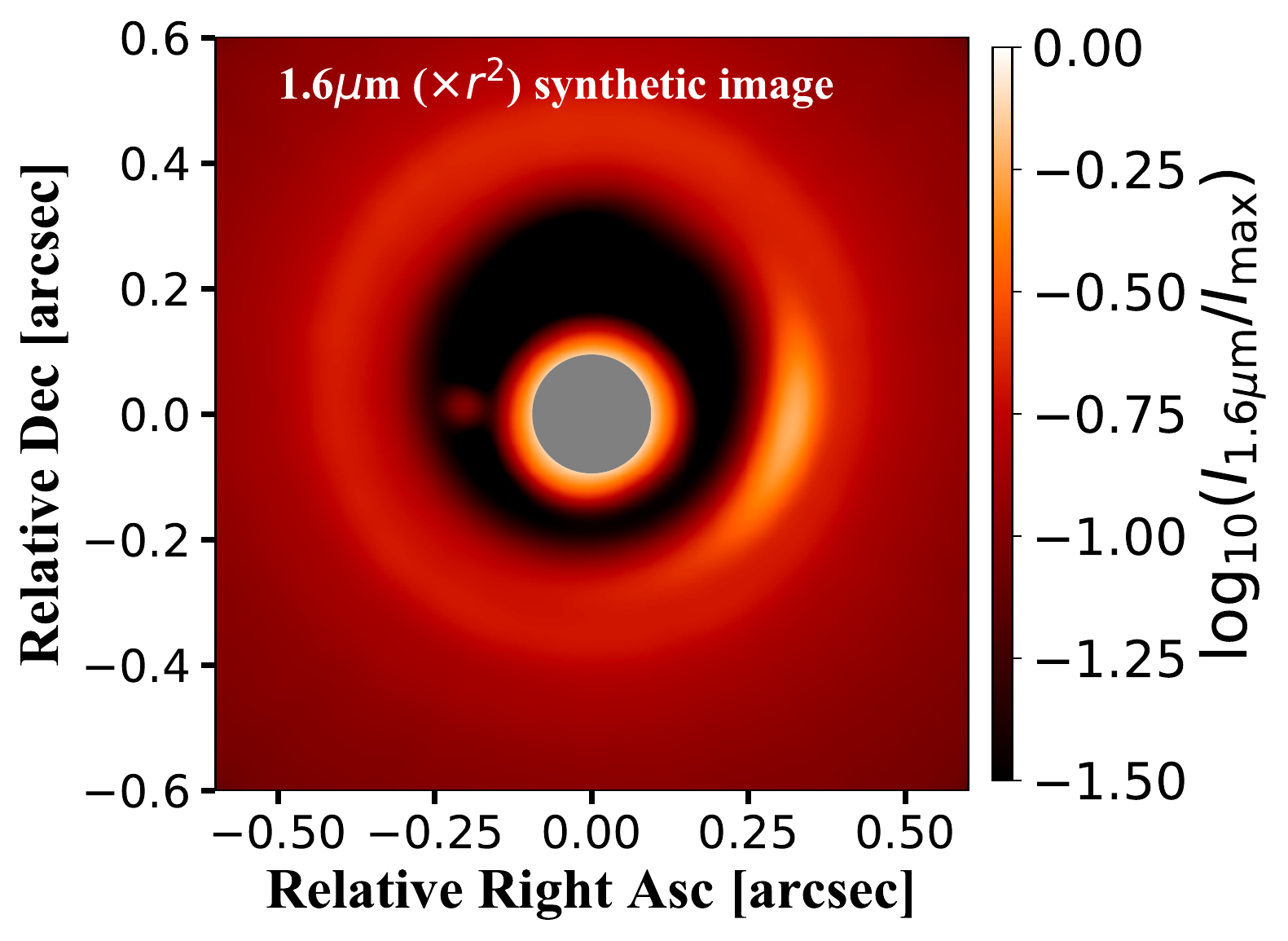}
    \caption{Synthetic image in H-band after convolving with a 0.04'' Gaussian beam and multiplying by $r^2$ using the model of a planet at 60\,au and an eccentricity of 0.1.}
    \label{fig:syntheticH}
\end{figure}

Figure~\ref{fig:syntheticH} shows the synthetic image (already deprojected) in the H-band from our radiative transfer calculations, after convolving with a 0.04'' Gaussian beam and multiplying by $r^2$. For this image, we use the  model of a planet at 60\,au and an eccentricity of 0.1 after 500 orbits.

\FloatBarrier

\end{appendix}
\end{document}